\documentclass[10pt,journal,compsoc]{IEEEtran}

\ifCLASSOPTIONcompsoc
  \usepackage[nocompress]{cite}
\else
  \usepackage{cite}
\fi

\usepackage{todonotes}

\usepackage{tikz}
\usepackage{xcolor}

\usepackage{booktabs, multirow} 
\usepackage{soul}
\usepackage{hyperref}

\usepackage{listings}
\usepackage{algorithm}
\usepackage{algorithmic}

\usepackage{amsmath}

\usepackage{todonotes}
\usepackage{xparse}
\newcounter{verbtodo}
\renewcommand{\theverbtodo}{\roman{verbtodo}}
\makeatletter
\NewDocumentCommand{\verbtodo}{v}{%
  \refstepcounter{verbtodo}\label{verbtodo@\theverbtodo}%
  \global\@namedef{verbtodo@\theverbtodo}{#1}%
  \addtocontents{tdo}{\defineverbtodo{\theverbtodo}|#1|}%
  \todo[inline, color=green!40]{\texttt{\expandafter\protect\csname verbtodo@\theverbtodo\endcsname}}%
}
\NewDocumentCommand{\defineverbtodo}{mv}{%
  \@namedef{verbtodo@#1}{#2}}
\makeatother

\usepackage{enumitem}
\usepackage{comment}

\usepackage[printonlyused]{acronym}
\newacro{isa}  [ISA]  {Instruction Set Architecture}
\newacro{sca}  [SCA]  {Side-Channel Analysis}
\newacro{ise}  [ISE]  {Instruction Set Extension}
\newacro{dma}  [DMA]  {Direct Memory Access}
\newacro{soc}  [SoC]  {System-on-Chip}
\newacro{sla}  [SLA]  {Side-channel Leakage Assessment}
\newacro{rtl}  [RTL]  {Register-transfer Level}
\newacro{tvla} [TVLA] {Test Vector Leakage Assessment}
\newacro{lti}  [LTI]  {Leakage Time Interval}
\newacro{lif}  [LIF]  {Leakage Impact Factor}
\newacro{cpa}  [CPA]  {Correlation Power Analysis}

\begin{document}

\title{Gate-Level Side-Channel Leakage Assessment with Architecture Correlation Analysis}

\author{Pantea~Kiaei,~\IEEEmembership{Student Member,~IEEE,}
        Yuan~Yao,
        Zhenyuan~Liu,~\IEEEmembership{Student Member,~IEEE,}\\
        Nicole~Fern,
        Cees-Bart~Breunesse,
        Jasper~Van~Woudenberg,
        Kate~Gillis,
        Alex~Dich,
        Peter~Grossmann,
        and~Patrick~Schaumont,~\IEEEmembership{Senior~Member,~IEEE}
\IEEEcompsocitemizethanks{
\IEEEcompsocthanksitem P. Kiaei, Z. Liu, and P. Schaumont are with the Department of Electrical and Computer Engineering, Worcester Polytechnique Institute, Worcester, MA, 01609.\protect\\
E-mail: \{pkiaei,zliu12,pschaumont\}@wpi.edu
\IEEEcompsocthanksitem Y. Yao was with the Bradley Department
of Electrical and Computer Engineering, Virginia Polytechnique Institute and State University, Blacksburg, VA 24061.\protect\\
E-mail: yuan9@vt.edu
\IEEEcompsocthanksitem N. Fern, C.-B. Breunesse, and J. Van Woudebberg are with Riscure North America, San Francisco, CA, 94108.
\IEEEcompsocthanksitem K. Gillis, A. Dich, and P. Grossmann are with Intrinsix Corp., Marlborough, MA, 01752.\protect\\
}
}

\IEEEtitleabstractindextext{%
\begin{abstract}
While side-channel leakage is traditionally evaluated from a fabricated chip, it is more time-efficient and cost-effective to do so during the design phase of the chip. We present a methodology to rank the gates of a design according to their contribution to the side-channel leakage of the chip. The methodology relies on logic synthesis, logic simulation, gate-level power estimation, and gate leakage assessment to compute a ranking. The ranking metric can be defined as a specific test by correlating gate-level activity with a leakage model, or else as a non-specific test by evaluating gate-level activity in response to distinct test vector groups. Our results show that only a minority of the gates in a design contribute most of the side-channel leakage. We demonstrate this property for several designs, including a hardware AES coprocessor and a cryptographic hardware/software interface in a five-stage pipelined RISC processor. 
\end{abstract}

}

\maketitle

\vspace{-4mm}
\section{Introduction}

Power-based side-channel leakage occurs when a secure chip performs operations that depend on an internal secret value such as a secret key. An adversary who observes the chip power consumption can derive the internal secret value through differential analysis techniques that correlate a power model of the secret activity with the observed power consumption.  In recent years, side-channel vulnerabilities have risen to prominence and successful side-channel attacks have been demonstrated on a wide range of devices from small IoT devices to large cloud computing systems. Therefore, the evaluation of the power-based side-channel leakage has become a critical component in the design flow of secure chips. It is particularly helpful to perform side-channel leakage assessment prior to manufacturing because it reduces the cost of post-manufacturing testing, and it reduces the probability of side-channel vulnerabilities in the chip tape-out.

\begin{figure}[t]
  \centering
  \includegraphics[width=\columnwidth]{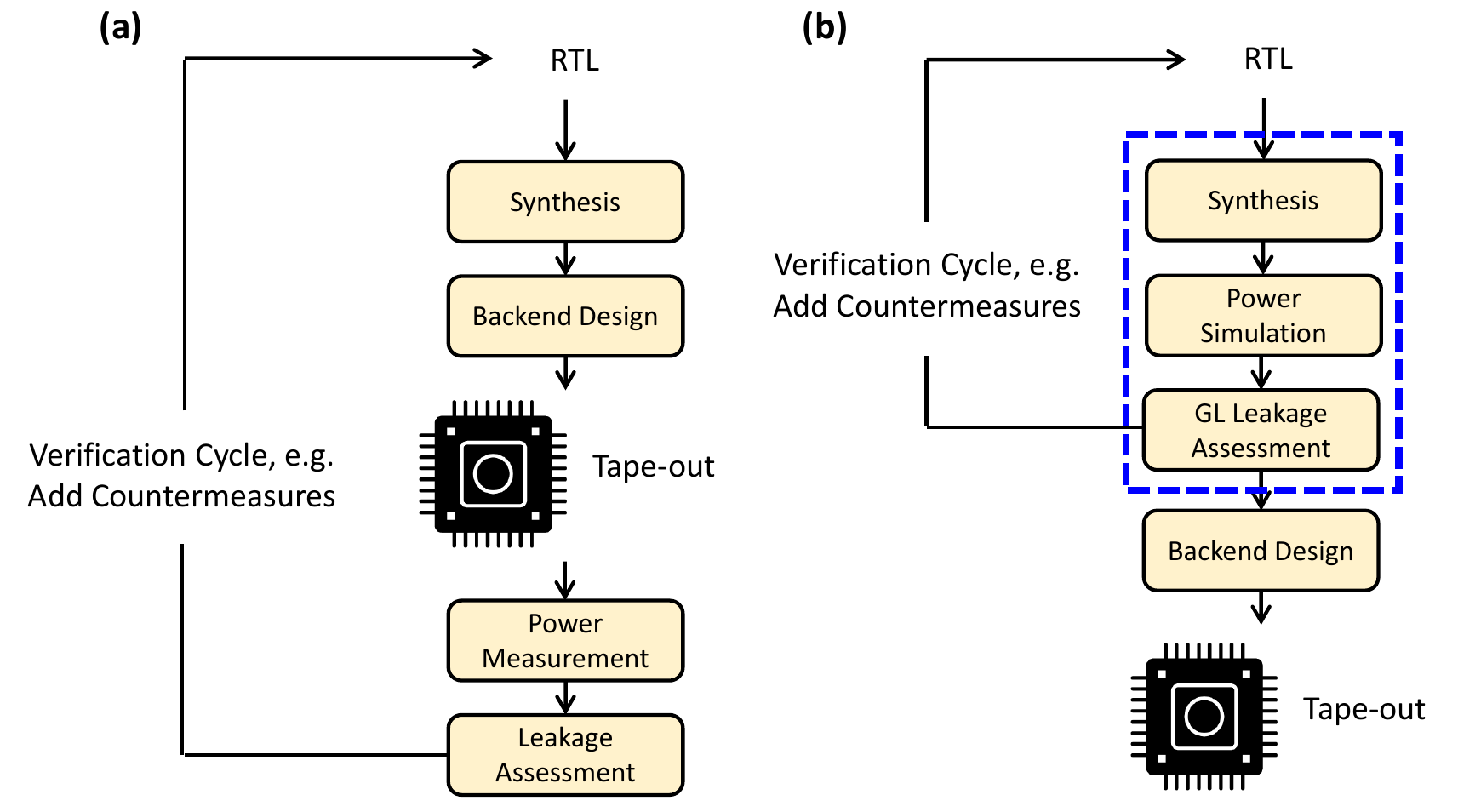}
  \caption{(a) Post-silicon side-channel leakage assessment flow. (b) Proposed flow. }
  \label{fig:intro}
 \end{figure}

\autoref{fig:intro} compares the \ac{sla} process of a post-silicon assessment flow with a pre-silicon assessment flow. The starting point is identical in both cases and assumes that a \ac{rtl} description of the design under consideration is available. With a post-silicon \ac{sla} flow, the \ac{rtl} design is first prototyped into a physical implementation. Power measurements are then collected from the design and statistically tested to confirm the presence of side-channel leakage or to estimate the quantity of side-channel leakage. In a pre-silicon strategy, the \ac{rtl} design is synthesized into a gate-level netlist including optional parasitic effects from place-and-route. Next, power traces are simulated and then statistically tested to confirm the presence of side-channel leakage.

These two flows appear similar from a macro-level objective, but they have very different properties. A post-silicon flow is expensive because of the extra prototyping step, which slows down the verification cycle. The statistical tests are applied globally on the measured leakage of the overall design. In a large and complex design, it therefore remains difficult to pinpoint the leakage source. 

In contrast, a pre-silicon flow makes use of simulated power traces, and it is able to perform side-channel leakage assessment at a fine granularity. In this paper, we present a side-channel leakage assessment methodology with a resolution of a single gate. The simulated traces of the pre-silicon flow are noiseless, and therefore they represent the attacker with the best possible observation. Due to the absence of noise, a pre-silicon flow can work with a fraction of the number of power traces compared to a post-silicon flow. On the downside, a pre-silicon flow must make a trade-off between accuracy and simulation speed. We use a gate-level power simulation methodology that is able to capture many technology-dependent effects (such as glitches \cite{nikova2006threshold} and static leakage \cite{DBLP:conf/ches/Moradi14}). Some side-channel leakage effects, including those based on coupling \cite{DBLP:conf/sersc-isa/ChenHS09} or the long-wire effect \cite{10.1145/3322483}, require a simulation accuracy beyond what gate-level power simulation can offer. Our proposed flow offers gate-level side-channel leakage assessment but makes no assertion of leakage below that abstraction level. 

This paper presents the following contributions. We describe a methodology called {\em Architecture Correlation Analysis} (ACA) which determines the side-channel leakage of a design at the granularity of a single gate.
The basic principle of ACA has been initially proposed in our previous work \cite{yao2020,yao2021pre}. This paper serves as an extension to our original publication. In this extended work,
we propose two different side-channel leakage assessment techniques for use in ACA. The first one is based on a specific test and it demonstrates the presence of correlation between a specific power model and individual logic gates. The second is based on a non-specific test that ranks a gate's power according to its ability to distinguish between two distinct groups of test vectors. The specific test is used to identify the gates that enable a specific side-channel attack, while the non-specific test is used to make a generic assessment on how much potentially harmful leakage can be produced by a gate. We apply our proposed ACA leakage assessment technique to two case studies: a cryptographic AES coprocessor, and the driver software for that coprocessor when running on a five-stage pipelined RISC processor. The side-channel leakage properties of AES are already well understood, and our experiments are specifically aimed at the ability of pre-silicon leakage assessment to identify the source of side-channel leakage. 

The remainder of the paper is organized as follows. The next section describes related work. Section 3 provides the overall methodology of Architecture Correlation Analysis (ACA), highlighting both the specific and the non-specific testing strategy. We also discuss a prototype implementation of the flow. Section 4 applies our proposed methodology to a cryptographic coprocessor. Section 5 applies the methodology to cryptographic driver software running on a RISC processor. Section 6 evaluates the performance of the proposed methodology. We then conclude the paper.
\vspace{-4mm}
\section{Related Work}

Gate-level side-channel leakage assessment is built on two components of design automation: (a) power simulation under a set of selected test vectors, and (b) identification of a leakage source at the sub-module level or gate level. We discuss related work on each of these two aspects.

\subsection{Power simulation for side-channel leakage analysis}

To simulate a design's power consumption, one needs a model of the design implementation details to estimate the power of the physical implementation under a set of test vectors. The model can be constructed at different abstraction levels, and there is a trade-off between modeling detail and simulation performance. Buhan {\em et al.} review many of the recent proposals to simulate side-channel leakage \cite{cryptoeprint:2021:497} and here we only describe the most representative ones. 


The origin of side-channel leakage power simulation is found in simulators for smart cards, starting with PINPAS \cite{PINPAS_2003}. The objective of these instruction-level simulations is to generate a power trace corresponding to a software application running on an embedded micro-controller. These simulators are processor-specific, and require knowledge of the internal design of the processor. Recent research efforts have addressed power model construction techniques to handle the case when the internal design is unknown. This includes ELMO \cite{ELMO_2017} and ROSITA \cite{ROSITA_2019} for power, and EMSIM \cite{EMSIM_2020} for Electromagnetic Radiation. In our approach, we build on the basic assumption that the hardware source code is available. 

It is now commonly understood that instruction-level power modeling by itself is inadequate to accurately capture all aspects of power-based side-channel leakage, and that additional modeling detail is required to capture circuit-level effects. The CASCADE power simulation flow aims at a comprehensive simulation of power traces at the gate-level \cite{CASCADE_2020}, while making the argument that gate-level power simulation hits a sweet spot for the known power-based side-channel leaks. A similar flow (and argument) is found in SCRIPT \cite{10.1145/3383445}. However, transistor-level power simulation has been investigated as well to address specific side-channel leakage assessments with a limited scope in time and in design size \cite{DBLP:conf/samos/RegazzoniBEGPDMPPLI07}. More recently, power simulation for side-channel analysis is starting to appear in commercial tooling. 

\vspace{-4mm}
\subsection{Identification of the leakage source}

By simulating power with a structural model, it becomes feasible to identify the {\em source component} of side-channel leakage. In traditional measurement-based side-channel leakage analysis, this type of analysis is not possible because the design under test remains a black box. We review several recent proposals aimed at identifying the structural source of side-channel leakage, starting at low abstraction levels. 

Karna partitions the gates of a design according to a spatial grid over the circuit layout \cite{karna2019}. A gate-level simulation leads to a power trace per grid cell. A leakage metric then ranks different cells according to their contribution to the side-channel leakage. The resolution of the Karna method depends on the granularity of the grid cells on the layout, since all logic cells within the same grid cell receive the same leakage score. The Karna authors include around 150 logic cells in one grid cell. Another tool, RTL-PSC. works at the register-transfer level of abstraction \cite{RTL_PSC_2019}. RTL-PSC analyzes the power in terms of transitions on the state variables of a design, while sub-cycle effects such as glitches and the effects of physical routing are abstracted out. The leakage metric ranks the different modules of a design. A similar register-transfer level analysis tool is PARAM \cite{DBLP:conf/host/FGBR20}, where the authors identify the sources of side-channel leakage in a processor's micro-architecture.

Another view on the problem of leakage source identification is to formally prove that a design meets a predefined side-channel leakage criteria. This technique works well for designs based on masking, a countermeasure based on secret-sharing. One example is Coco, which combines event-driven simulation with a SAT-solver-based verification of the statistical distribution of the secret shares \cite{DBLP:conf/uss/GigerlHPMB21}. Strictly speaking, these tools do not simulate the power consumption, but they verify the statistical properties of design activity.

Compared to this related work, ACA creates a ranking of the cells in the design according to their contribution to side-channel leakage, with a user-defined leakage metric. ACA can handle both specific and non-specific leakage testing. ACA is processor-independent as well as technology-independent. ACA builds a flow on commercially available synthesis and power simulation tooling. To the best of our knowledge, no such tool has been presented by earlier work.

\vspace{-4mm}
\section{Architecture Correlation Analysis}

In this section, we outline the strategy of Architecture Correlation Analysis. The objective of ACA is to identify a ranking among the cells\footnote{We use the term {\em cell} over {\em gate} as it reflects better the technology encountered in standard-cell based IC design. A single cell often corresponds to multiple primary gates.} of a design according to their contribution to side-channel leakage. The cell ranking does not have to reflect the absolute level of side-channel leakage generated by a cell. Knowing just a relative ranking already provides critical insight into the parts of a design that are most prone to side-channel attacks. 

Traditional side-channel leakage assessment uses the overall power consumption of a design to make an assessment on the global design. In contrast, ACA uses not only the overall power consumption but also the internal design construction details to make an assessment of leakage on a {\em part} of a design and to rank these local assessments. ACA uses gate-level power simulation in order to capture power events with sub-cycle accuracy as well as structural effects such as wire-loading and static leakage. The challenge of ACA is to perform such a gate-level side-channel leakage assessment with reasonable accuracy but {\em without} exhaustively generating the power consumption trace for each individual cell in the design. 

Side-channel leakage assessment aims to minimize assumptions regarding the specific strength and know-how of the attacker. This leads to the use of specific and non-specific tests. A {\em specific} test for side-channel leakage uses a high-level power model that the adversary would presumably use in a Differential Power Analysis. A {\em non-specific} test uses two groups of inputs and aims to demonstrate a statistically distinguishable power consumption difference between those two groups. The \ac{tvla} methodology provides guidance on the selection of the two groups of input vectors \cite{gilbert2011testing}. Both specific and non-specific tests have their merits and limitations. Specific tests are limited by specific assumptions on the capabilities and activities of the attacker, but provide specific assertions on the existence of a side-channel attack. Non-specific tests avoid such assumptions, but they are unable to assert the existence of a side-channel attack that can exploit the leakage. We, therefore, present an ACA methodology for either approach.

\vspace{-4mm}
\subsection{Overall Methodology}

The ACA methodology includes three phases: (a) activity trace and power trace generation, (b) leakage time interval selection, and (c) leakage impact factor evaluation. The first phase is common to specific and non-specific tests, while the second and third phases differ according to the testing strategy. We will discuss each phase separately.

\begin{figure}[t]
  \centering
  \includegraphics[width=\columnwidth]{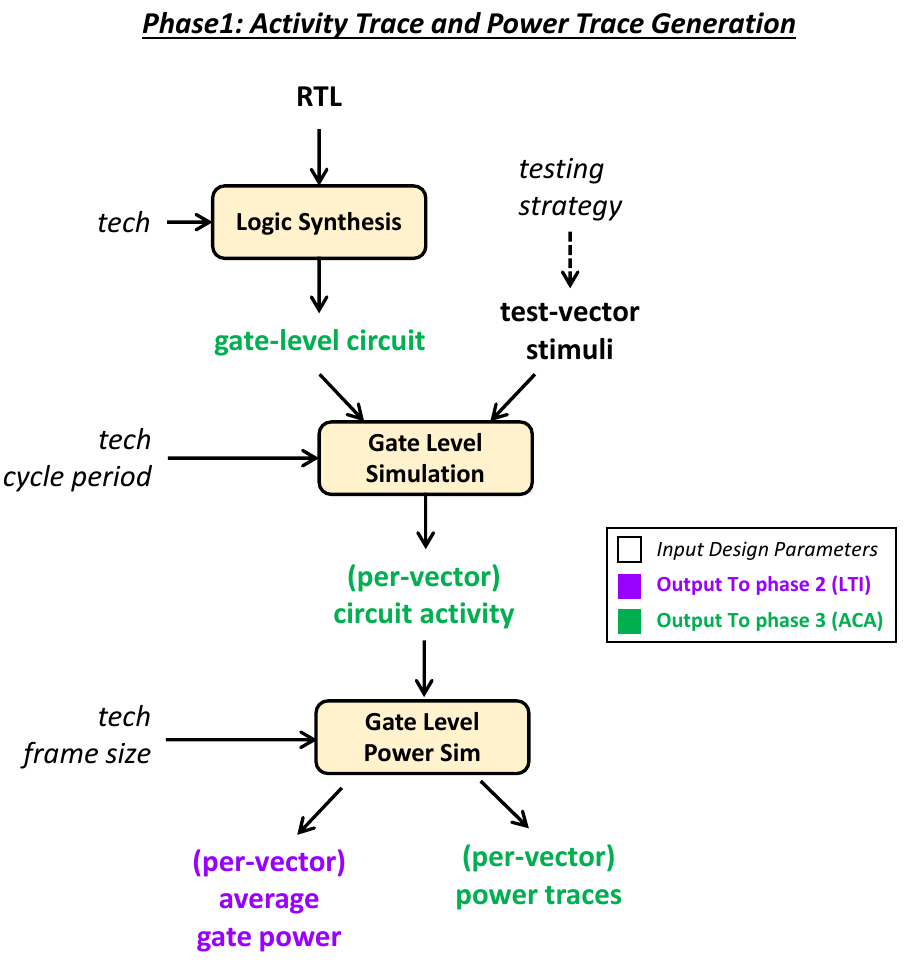}
  \caption{ACA Phase 1: Stimuli and Trace Generation for ACA}
  \label{fig:phase1}
 \end{figure}

\autoref{fig:phase1} describes the common first phase of specific and non-specific ACA, covering logic synthesis, gate-level simulation, and gate-level power simulation. The design parameters, shown in italic, include the testing strategy, the target technology, the target cycle period, and the frame size. The frame size is the time step used in the traces of the power simulator. All of the intermediate results of the flow are used by later phases of ACA.

Logic synthesis transforms the input RTL under a given performance constraint (speed/area) into a gate-level netlist. Next, the gate-level netlist is simulated for a set of test vectors while recording the circuit activity for each net in the design over the simulation time window of interest. The type and number of test vector stimuli depend on the assessment type (specific/non-specific) and the acceptable statistical uncertainty of the leakage assessment result. We will address the selection of test vectors in the next subsections. 

Two factors greatly help in reducing the number of test vectors for a side-channel leakage assessment. The first factor is that power simulation is noiseless so that each test vector and internal design state needs to be simulated only once. The second factor is that each simulation run can be isolated from the next by re-initializing the design in between simulation runs. In a typical side-channel leakage assessment, we gather  between a few hundred and a few thousand simulated traces and we aim for a turn-around of all three phases of ACA in less than 24 hours.

The gate-level power simulation estimates the time-varying power consumption of the design for each test vector as follows. First, the simulation time window is partitioned into frames. Next, for each frame, the gate-level power simulator computes the average power of a gate as a combination of the switching power, the internal power, and the leakage power. The gate switching power depends on the per-frame output toggle rate and the capacitive loading at the gate output. The gate internal power depends on the per-frame, per-pin input toggle rate. The gate leakage power depends on the per-frame state-dependent leakage power. All these factors are scaled by the technology selection and by the gate type. The sum of all these factors for all active gates within the frame determines the average frame power.

\begin{figure}[t]
  \centering
  \includegraphics[width=\columnwidth]{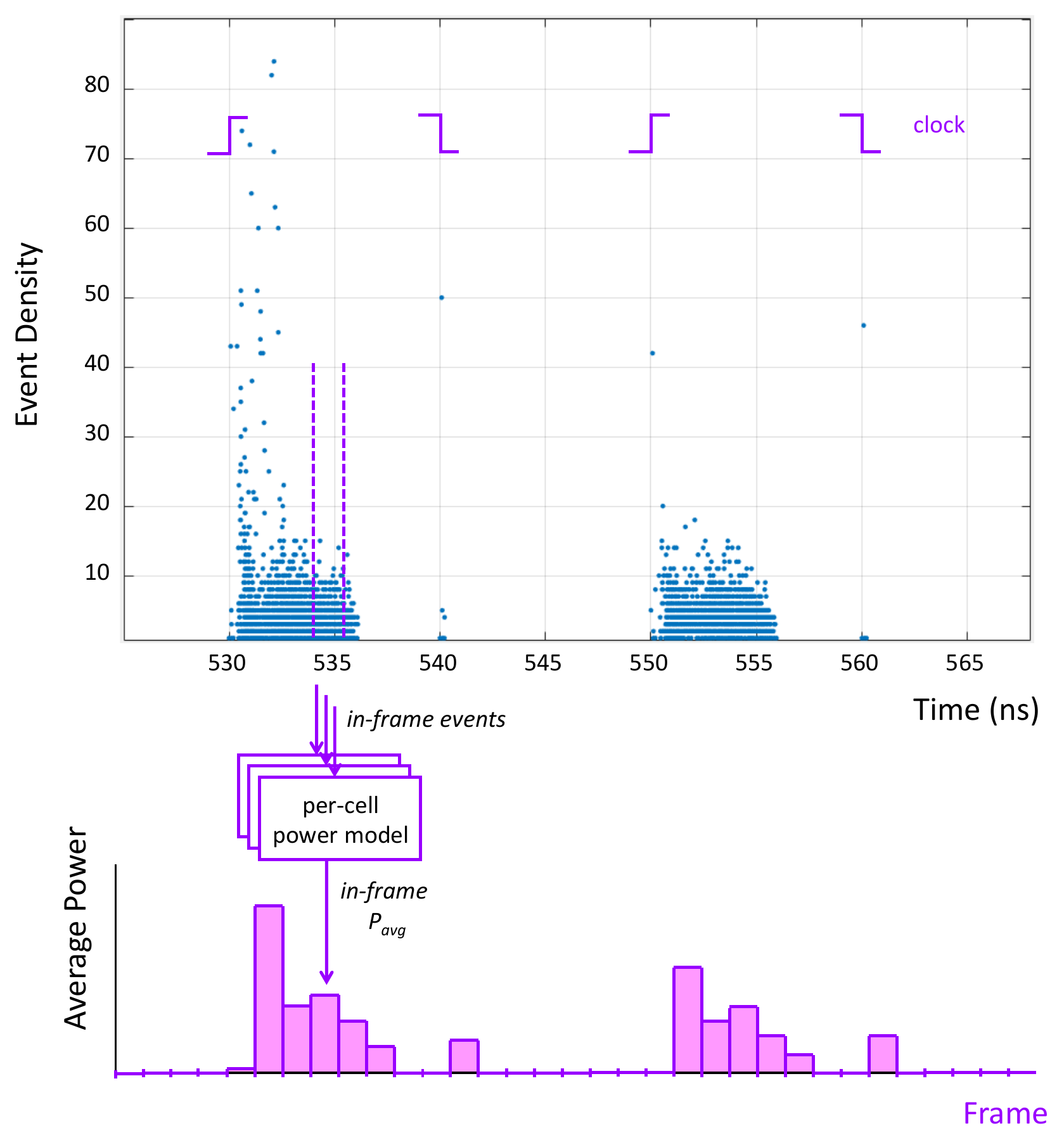}
  \caption{Event Density for two cycles of an 9,640-cell hardware AES design}
  \label{fig:events}
 \end{figure}

The frame size is an important selection parameter of the gate-level power simulation. The frame size determines the smallest time interval analyzed by the ACA flow for side-channel leakage. Each single cell in a design typically switches over a time interval much smaller than the clock cycle, and much smaller than the frame size. A single frame will thus contain the leakage of many different cells. A smaller frame size helps in detecting power variations caused by a specific cell. \autoref{fig:events} illustrates this point in further detail. The top half of the figure is an event density plot for two cycles from a hardware AES design containing 9,640 cells. The cycle time of this design is 10ns, and after each up-going clock edge, there is about 5ns of activity as the combinational cells settle to the new register output. In this simulation, there are about 10,000 events in the first clock cycle, and 6,500 events in the second clock cycle. To estimate the power, we select 8 frames per clock cycle (as an example). The power simulator will then compute the power per frame by analyzing the events within that frame. The power trace will thus contain 8 points per clock cycle, and side-channel leakage must be detected by power variations on any of these points. Clearly, with a smaller frame size, fewer events will contribute to that frame, so that it becomes easier to identify which events (and which cells) are a root cause of side-channel leakage.

On the other hand, a {\em large} frame size is beneficial because it improves simulation time.  We demonstrate this effect with the following experiment. A design containing 1, 4 resp. 16 AES S-boxes is driven by a set of counters that each count from 0 to 255. We determine the power estimation cost for a 256-cycle test-bench under various frame sizes. The total simulated time remains 256 cycles for each case, and therefore a smaller frame size requires more frames to be computed. \autoref{tab:framesize} demonstrates that the power simulation cost significantly depends not only on the design size, but also on the number of frames. 

In a practical assessment of hardware, we over-sample the clock cycle at least several times, in order to run the analysis on sub-cycle events. However, with long-running, complex simulations, we have already successfully down-sampled the frame size to as much as 80 clock cycles per frame, while still being able to demonstrate side-channel leaks \cite{cryptoeprint:2021:1235}.
Successfully finding side-channel leaks in an under-sampled power trace is due to the averaged sampling technique employed by power simulation tools \cite{kiaei2022leverage}.

\begin{table}[!t]\centering
\caption{Normalized complexity of power estimation time for three different design sizes and four different frame widths.}\label{tab:framesize}
\vspace{-.4cm}
\scriptsize
\begin{tabular}{lrrrr}\toprule
\textbf{Samples/cycle} &  \textbf{\# Frames} & \textbf{1 SBOX} & \textbf{4 SBOX} & \textbf{16 SBOX} \\\midrule
1/256           & 1      & 1.0     &  1.72     &   4.54   \\
1               & 256    & 6.9     &  23.5     &   93.8   \\
2               & 512    & 11.1    &  37.1     &   149.4  \\
4               & 1000   & 13.17   &  45.6     &   184.8  \\
\bottomrule
\multicolumn{5}{l}{Skywater 130nm power simulation with Cadence Joules}
\end{tabular}
\vspace{-.5cm}
\end{table}

In Phase 2 and Phase 3 of the ACA flow, we aim to identify the cells' individual contribution to side-channel leakage. The challenge is to complete this task using only the global power traces. Indeed, the generation of per-cell power traces has quadratic complexity (namely {\em $design\_size \times frame\_count$}) and is therefore not scalable to large applications. We solve this with a two-step approach. First, using the global power traces, we identify the {\em leakage time interval}, the time window within which a design leaks information. Next, within the leakage time interval, we use the activity traces to identify the contribution of an individual cell to a design-level leak. Therefore, Phase 2 and Phase 3 of the ACA flow are defined as {\em Leakage Time Interval Selection} and {\em Leakage Impact Factor Computation} respectively. Leakage Time Interval Selection uses design-global power traces, while Leakage Impact Factor computation uses per-cell event traces.

To test individual cells for side-channel leakage, we can use two different testing scenarios. In the following sections, we clarify each testing scenario separately.

\vspace{-4mm}
\subsection{ACA for Specific Testing}

\begin{figure}[t]
  \centering
  \includegraphics[width=\columnwidth]{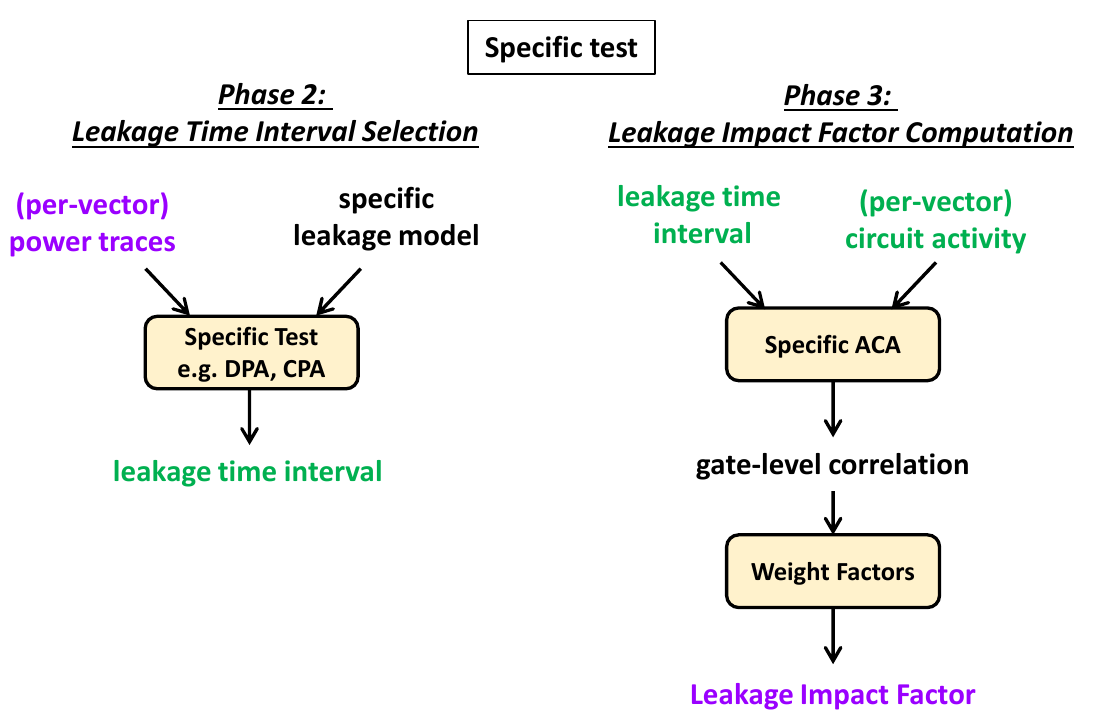}
  \caption{Phase 2 (Leakage Time Interval) and Phase 3 (Leakage Impact Factor) computation for specific ACA}
  \label{fig:specific}
 \end{figure}

\autoref{fig:specific} describes the two steps to apply ACA using a specific test. The specific test requires the definition of a leakage model, similar to the leakage model used in Differential Power Analysis or Correlation Power Analysis. The leakage model is correlated with the simulated power consumption to identify a window of high correlation as the \ac{lti}, the time window during which a design leaks. Next, each individual cell is ranked by computing its correlation to the leakage model within the \ac{lti}.

\subsubsection{Leakage Time Interval for Specific Testing}

Given a cipher which computes an internal result $V_k = f(K, C_k)$ with a partial key $K$ and a controlled input $C_k$ from $k$-th test vector, a possible leakage model $L(V_k)$ is the Hamming Weight $HW(V_k)$. Alternately, for an internal tuple $(V1_k,V2_k) = f(K, C_k)$, the leakage model can be expressed as the Hamming Distance $HD(V1_k,V2_k)$. We then compute the correlation between the leakage model $L(V_k)$ or $L(V1_k,V2_k)$ and the simulated power $P_k[n]$ for each frame $n$ of the simulated power traces:

\begin{equation}
    \rho[n] = \frac{cov(L(V_k), P_{k}[n])}{\sigma_L \sigma_P}
\end{equation}

The \ac{lti} consists of the set of frames for which the absolute correlation is above a given threshold.

\begin{equation}
    LTI = \{n\} : abs(\rho[n]) > \rho_{threshold}
    \label{eq:lti}
\end{equation}

The choice of the threshold is a sensitivity parameter that must be chosen such that the \ac{lti} covers design activity that contains a likely correlation peak. \autoref{tab:lif} shows several examples of threshold levels as a function of the number of test vectors $m$ and the confidence level. As expected, a requirement for higher confidence or the use of fewer traces will increase the confidence interval, which means that stronger correlation peaks must be identified before the frame is flagged as leaky and added to the \ac{lti}.

\begin{table}[t]
\centering
\caption{Pearson Correlation Threshold Levels as a function of test vectors $m$ and confidence}
\label{tab:lif}
\vspace{-.4cm}
\scriptsize
\begin{tabular}{cccc}\toprule
\textbf{Confidence Level} & \textbf{m=600} & \textbf{m=1000} & \textbf{m=2000}\\\midrule 
99\%       & $\pm$ 0.105  & $\pm$ 0.081  & $\pm$ 0.058  \\ 
95\%       & $\pm$ 0.080  & $\pm$ 0.062  & $\pm$ 0.044  \\
90\%       & $\pm$ 0.067  & $\pm$ 0.052  & $\pm$ 0.037  \\
\bottomrule
\end{tabular}
\vspace{-.5cm}
\end{table}

\subsubsection{Leakage Impact Factor for Specific Testing}

Within the \ac{lti}, we  compute the contribution of each individual cell to the leakage. This contribution is quantified in the \ac{lif}, which is a dimensionless number that expresses the relative amount of side-channel leakage from a cell.

\begin{figure}[t]
  \centering
  \includegraphics[width=\columnwidth]{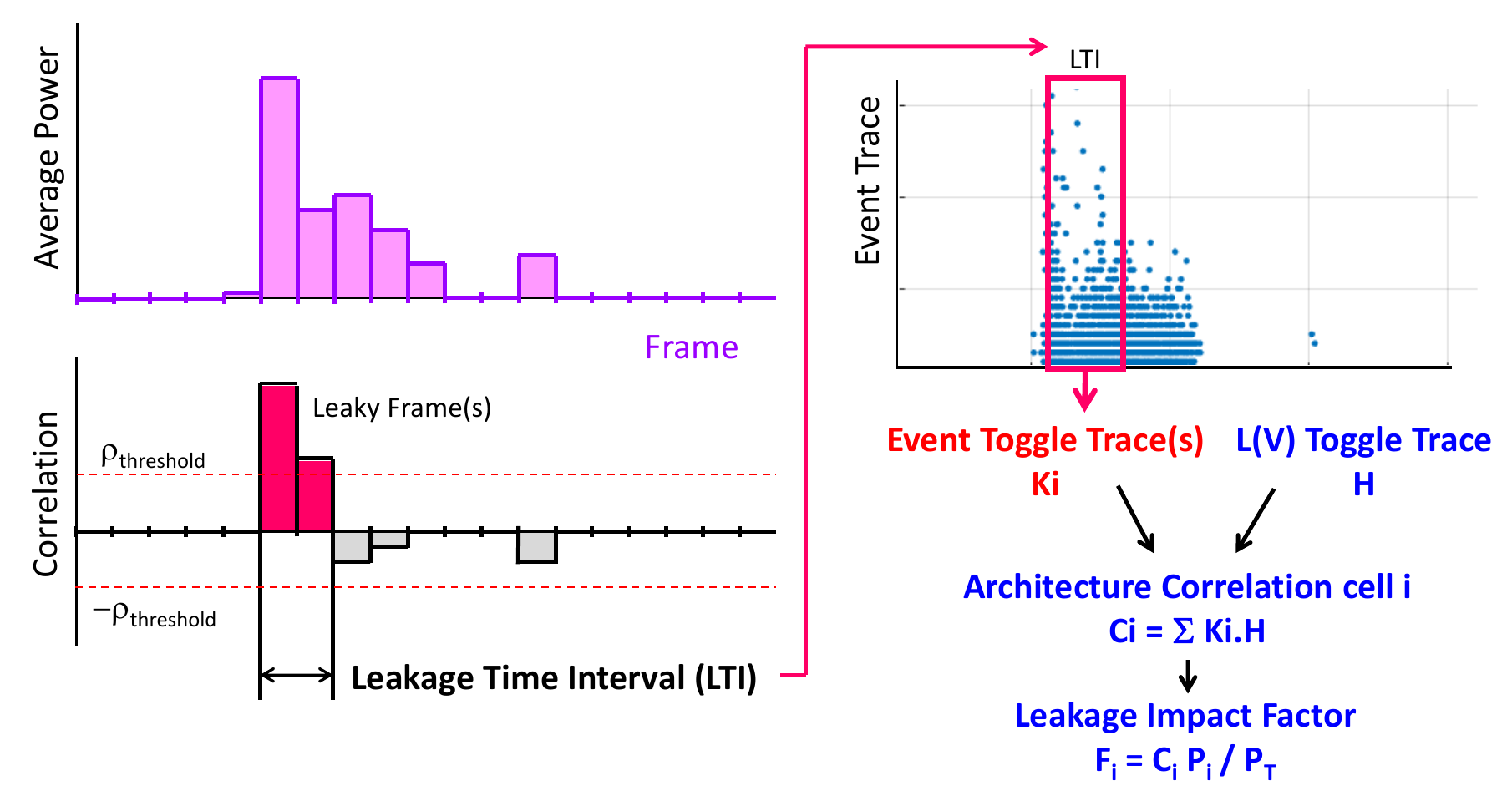}
  \caption{(left) Leakage Time Interval Detection (right) Architecture Correlation Analysis}
  \label{fig:lifcomp}
 \end{figure}

The \ac{lif} of a cell is computed as the correlation of cell output activity and leakage model activity. The  \ac{lti} is a set of frames that are considered {\em leaky} (\autoref{fig:lifcomp}, left). To investigate the cell leakage, the \ac{lti} is superimposed over the activity trace. For each net (or each net-driving cell), we then compute a toggle trace as follows. When a given net switches during the \ac{lti}, then that transition is counted as a +1 toggle. When a given net does not include such a transition in the \ac{lti}, then that is counted as a -1 toggle. Hence, the activity of each net $i$ under test vector $j$ is converted to a bi-valued signal $K_{ij}$ with values \{-1, +1\}. To compute the architecture correlation $C_i$ of net $i$, $K_{ij}$ is multiplied with the toggle trace $H_j$ of the leakage model $L(V)$ (\autoref{fig:lifcomp}, right).

\begin{equation}
    C_i = \sum_{stimuli} K_{ij} . H_j
\end{equation}

This correlation can be computed for every frame within the \ac{lti}. A high value in $C_i$ indicates a strong correlation between the cell activity and the leakage model, and hence a strong indication that the cell contributes side-channel leakage. All cells in the design are ranked according to their $C_i$ from most leaky to least leaky. We also include an additional weighing factor for each $C_i$, defined as the ratio of the cells' average power consumption $P_i$ during the \ac{lti} over the total average power consumption of all gates $P_T$. This increases the weight of high-drive cells with a high correlation. This leads to the weighed per-cell \ac{lif} $F_i$:

\begin{equation}
    \label{eq:weighing}
    F_i = C_i \frac{P_i}{P_T}
\end{equation}

Unlike computing a gates' power for every frame, computing a gates' average power over all frames is relatively quick (\autoref{tab:framesize}).
Therefore, the weighing process is scaleable. Each \ac{lif} factor is bound to a specific cell within a specific frame in the \ac{lti}. Hence, for a design with $G$ gates and $J$ leaky frames, the list of \ac{lif} factors contains $G \times J$ entries. These entries are sorted by \ac{lif} value to determine the overall leakage ranking.

\vspace{-3mm}
\subsection{ACA for Non-specific Testing}

\begin{figure}[t]
  \centering
  \includegraphics[width=\columnwidth]{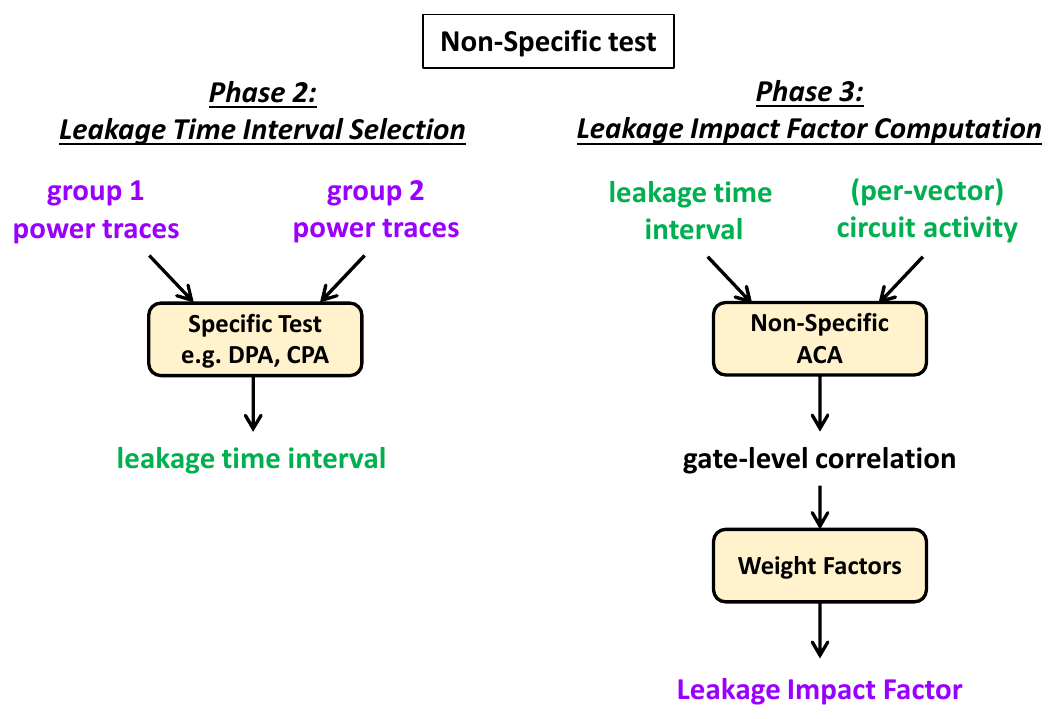}
  \caption{Phase 2 (Leakage Time Interval) and Phase 3 (Leakage Impact Factor) computation for nonspecific ACA}
  \label{fig:nonspecific}
 \end{figure}

When a specific test is hard to apply, or when the number of leakage models $L(V)$ becomes too numerous, it may be helpful to apply a more generic non-specific test for leakage. We can run ACA using a non-specific leakage model following a strategy as in \autoref{fig:nonspecific}. Similar to \ac{tvla}, non-specific ACA requires the definition of two groups of stimuli. These two groups are should exhibit some systematic difference in the design behavior. For example, Goodwill {\em et al.} suggest AES test vectors that are random for group 1, while introducing a specific bias within a middle-round state for group 2. Using these two vector groups, ACA then follows the same two-step strategy as for specific testing. First, the \ac{lti} is computed to bound the leakage in time, and next the \ac{lif} per gate is computed. The testing statistic is  adjusted to a non-specific test.

\subsubsection{Leakage Time Interval for Non-Specific Testing}

The test-statistic compares the distribution of power values at a specific frame between two test vector groups. One solution is to use a Welch-t statistic, which tests the difference between the mean values of both groups.
Another solution is to measure the correlation of the power value to the group number. We use the leakage model $N = groupid$, with $groupid$ equal to -1 for vectors from group 1, and +1 for vectors from group 2. With this leakage model, we can compute a non-specific test statistic as a correlation value that can be compared against a threshold value  $\rho_{threshold}$. With $P[n]$ the power consumed during frame $n$, the correlation is given by:

\begin{equation}
    \rho[n] = \frac{cov(N, P[n])}{\sigma_N \sigma_P}
\end{equation}

The \ac{lti} is then defined using the same method as in \autoref{eq:lti}.

\subsubsection{Leakage Impact Factor for Non-Specific Testing}

Once the \ac{lti} is fixed, we proceed with computing the \ac{lif} using a similar strategy as for the specific test. The \ac{lti} is superimposed over the activity trace of the design. A net transition during the \ac{lti} counts as a +1 toggle, and an absence of transition counts as a -1 toggle. We compute the architecture correlation $C_i$ of net $i$ by correlating the toggle trace $K_{ij}$ with the groupid $N_j$. 

\begin{equation}
    C_i = \sum_{stimuli} K_{ij} . N_j
\end{equation}

Again, a high value in Ci indicates a strong correlation between the cell activity and the non-specific leakage model, and hence flags the cell $C_i$ as leaky. To find a cell's  non-specific leakage impact factor $F_i$, a weighing factor is introduced as in \autoref{eq:weighing}.

The advantage of the non-specific test over the specific test is that no high-level leakage model is necessary. For example, we have used non-specific tests based on one or more state bytes at a middle round being 0 or else random. In the experimental results, we will demonstrate the selectivity of both the specific as well as the non-specific ACA method.

\vspace{-4mm}
\subsection{Implementation}

Our flow is fully realized in commercial tooling along with customized scripting to implement the statistical post-processing. We use Cadence Genus 20.1 for logic synthesis from RTL, Cadence XCelium 20.09 for functional simulation, and Cadence Joules 10.1 for gate-level power simulation. Computation of ACA \ac{lif} and \ac{lti} is scripted on top of the Jlsca toolbox\footnote{\url{https://github.com/Riscure/Jlsca}}.
We have used Skywater 130nm standard cells as technology targets during experiments.

\vspace{-4mm}
\section{ACA on a Cryptographic Coprocessor}
\label{sec:coproc}

\begin{figure}[t]
  \centering
  \includegraphics[width=0.8\columnwidth]{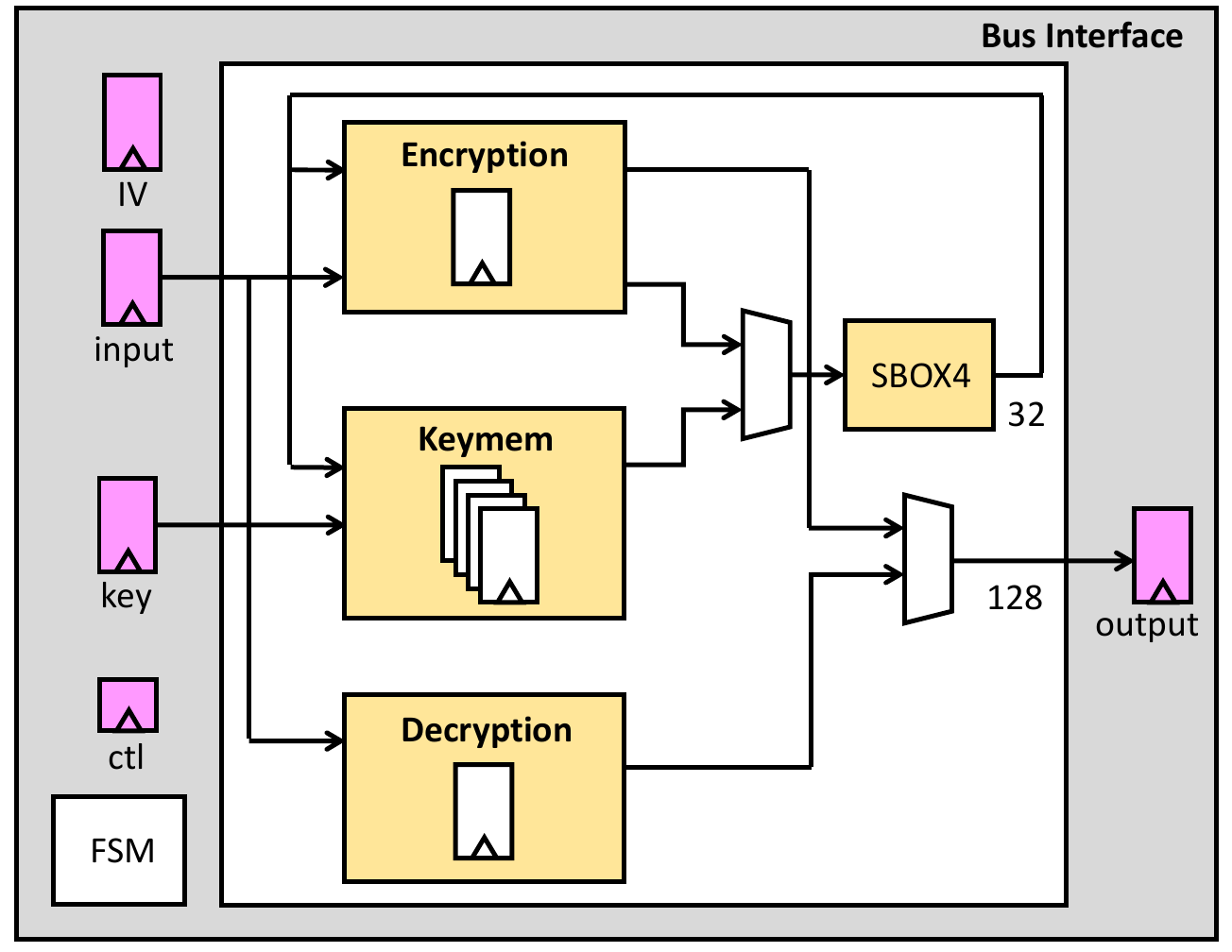}
  \caption{Block diagram of the AES encryption/decryption unit}
  \label{fig:block}
\end{figure}

In this section, we describe the application of ACA on an AES encryption/decryption coprocessor. The  architecture selected for analysis is typical for a medium-throughput accelerator residing in an embedded SoC. The coprocessor handles encryption and decryption and uses an offline key schedule, which computes the roundkeys once upon loading of the key. A single round takes 5 clock cycles. In the first four cycles, the coprocessor computes 16 Sbox lookups in sets of 4, and in the fifth clock cycle, the remainder of the round is computed. The encryption/decryption core is encapsulated by a bus interface which contains software-accessible registers and a controller. The bus interface handles various modes of operation for the coprocessor. We synthesized this coprocessor for SkyWater 130nm standard cell technology and a 50MHz clock. \autoref{tab:picoaes} reports the type and number of cells used in the design, as well as their relative area.

\begin{table}[t]
\centering
\caption{Cell type and area for AES coprocessor}
\label{tab:picoaes}
\vspace{-.4cm}
\scriptsize
\begin{tabular}{lrr}\toprule
\textbf{Type} & \textbf{Cell Count} & \textbf{Area (\%)} \\\midrule 
Sequential         & 2,479 & 51.8 \\
Logic              & 7,161 & 48.2 \\
Total              & 9,640 & 100.0  \\
\bottomrule
\end{tabular}
\vspace{-.5cm}
\end{table}

\vspace{-4mm}
\subsection{Architecture Correlation Analysis}

Stimuli selection plays an important role in ACA, as it enables a designer to choose which part of a design will be exercised. In this analysis, we will focus on Architecture Correlation Analysis for a single key byte in the AES coprocessor. The objective of the ACA is to determine which cells, among the 9,640 cells in the design, contribute to side-channel leakage of this key byte. We explicitly differentiate this objective (finding leaky gates) from a more traditional side-channel analysis of the hardware. There is no doubt that there is side-channel leakage in this design. However, the object of this experiment is to find {\em what cells are most responsible for this leakage}?

\begin{figure}[t]
  \centering
  \includegraphics[width=\columnwidth]{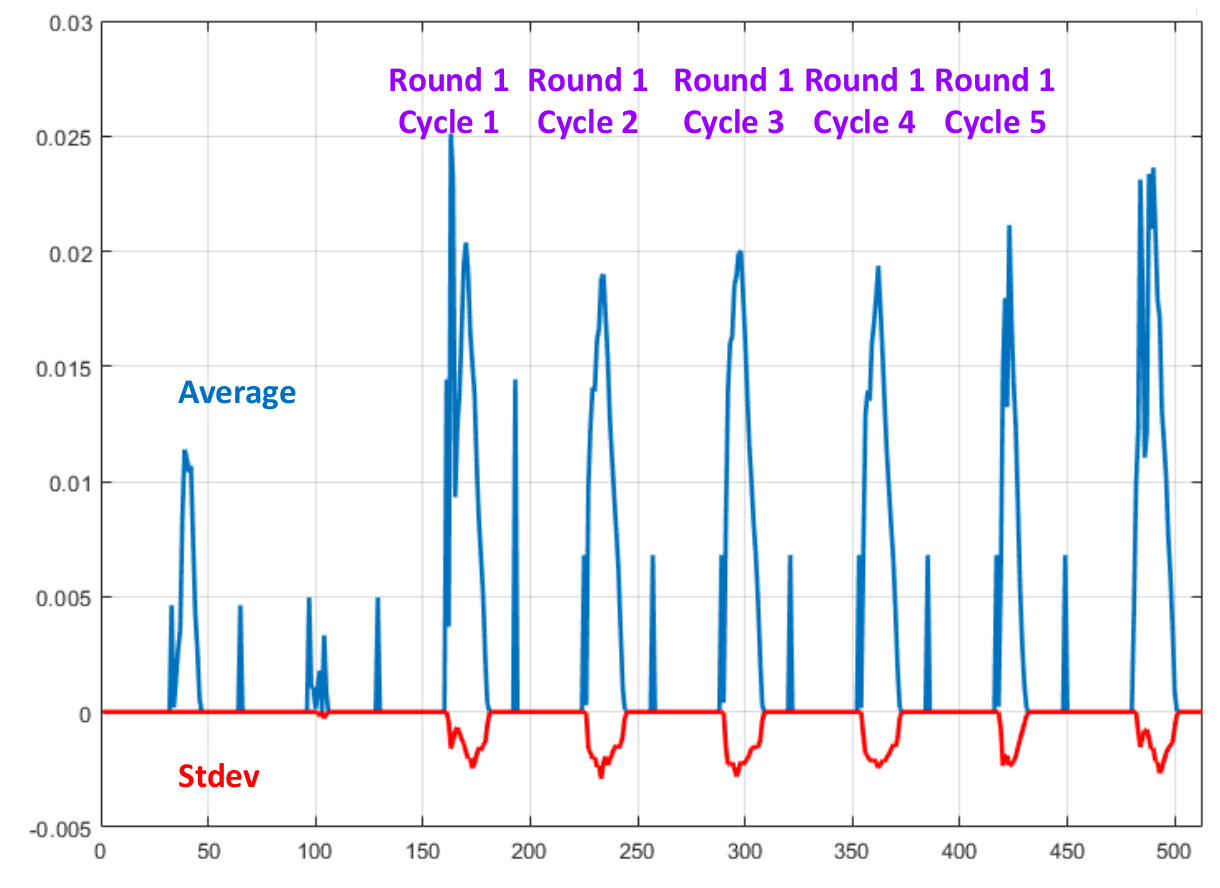}
  \caption{Average Power Trace and Standard Deviation of AES Coprocessor in first round}
  \label{fig:picopower}
\end{figure}

\subsubsection{Results}

We performed ACA as follows. We selected a set of 1024 vectors under a random plaintext and a constant key. Next, we ran a gate-level simulation and a gate-level power simulation for ACA. \autoref{fig:picopower} shows the average power trace of the first round at 64 frames per clock cycle, as well as the (sign-flipped) standard deviation. The clarity of this power trace illustrates the strength of noiseless simulation and outlines each clock cycle of operation as well as the location of power variations.  In this simulation, there are 512 frames in each power trace.

We next ran ACA using a specific leakage model on the Hamming Weight of the SBOX output of the first key byte. We selected a specific leakage model on the Hamming Weight of the SBOX output of the first state byte. We identify the \ac{lti} with a $\rho_{threshold}$ of 0.2, which flags 230 frames out of the 512 frames as containing potentially leaky samples.  By correlating the transitions by cells within \ac{lti} with the specific leakage models, 412 cells are then flagged as leaky. This group represents 4.3 \% of the total number of 9,640 cells. We will analyze the relation of these 412 cells to the overall AES coprocessor in \autoref{sec:lga1}.

\subsubsection{Result Verification}

\begin{figure}[t]
  \centering
  \includegraphics[width=\columnwidth]{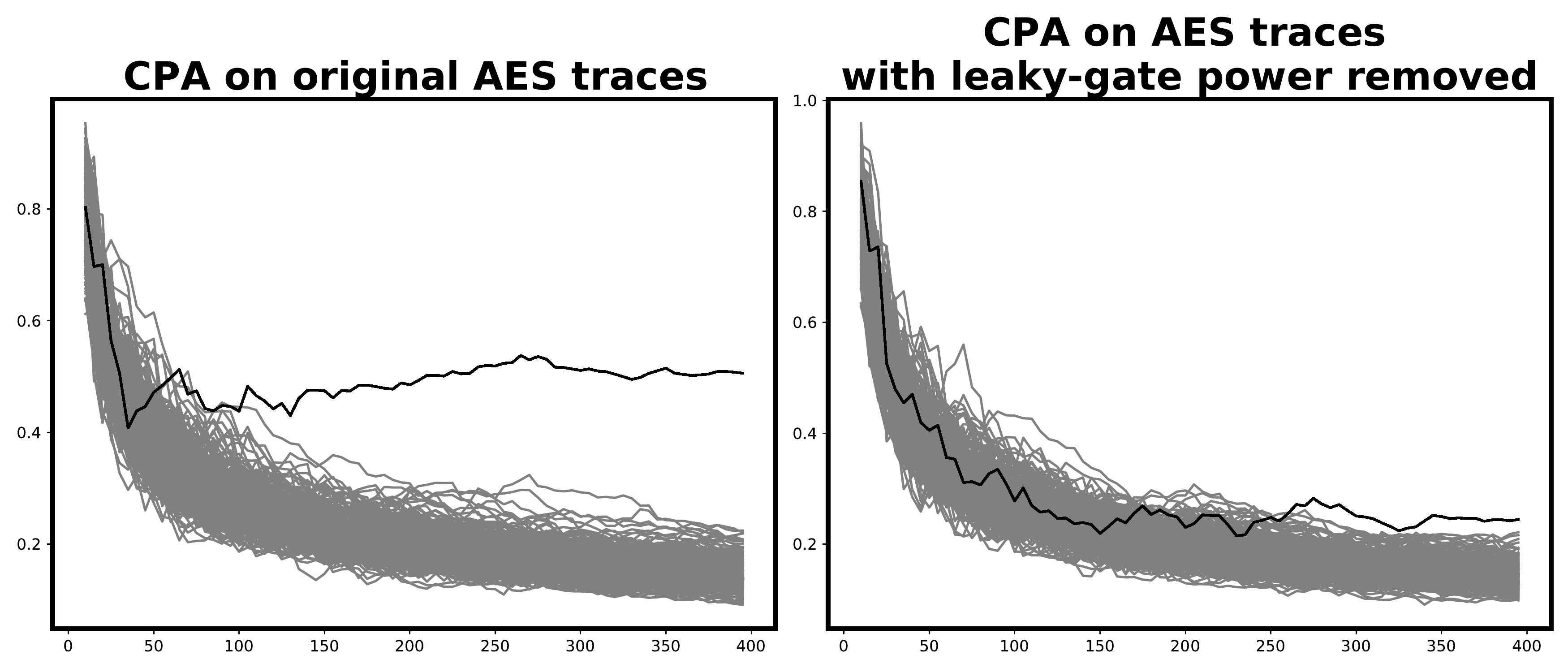}
  \caption{(left) CPA on AES Coprocessor traces reveals a correlation peak at about 75 traces (right) CPA on modified AES Coprocessor traces significantly delays correlation peak disclosure to at least 250 traces.}
  \label{fig:picoleaky}
\end{figure}

The assertion made by ACA is that the side-channel leakage under the selected leakage model is primarily caused by these 412 cells. To verify the correctness of the selection, we performed the following verification. We re-ran the simulation while collecting individual power traces for {\em each} cell for all vectors and all frames. The per-cell power traces of the 412 selected leaky cells are then subtracted from the overall power traces to construct a modified set of power traces. Next, we apply a \ac{cpa} on the original trace set as well as on the modified trace set, with the results summarized in \autoref{fig:picoleaky}(left: unmodified set, right: modified set). For the modified set, the number of measurements to disclosure increases with a factor of 3.

We emphasize that this \ac{cpa} experiment only verifies the selection of leaky cells. ACA is not a countermeasure but a detection tool. One cannot remove an arbitrary cell from a netlist without substituting it with an equivalent cell with identical functionality. However, ACA is useful in conjunction with countermeasure tools that protect individual cells or subsets of cells, such as Karna \cite{karna2019} or STELLAR \cite{DBLP:conf/host/DasNCGS19}.
A second observation is that a power simulation that collects an individual power trace for every gate is extremely complex both in disk space and in time. We found the overhead of single-gate power tracing (compared to standard ACA) to be around 4 orders of magnitude in disk storage and one order of magnitude in simulation time, and worsening with design size. The high cost of per-gate power tracing highlights the strength of ACA to use per-gate activity traces, which are a byproduct of functional verification.

\subsubsection{Impact of Frame Size}

\begin{figure}[t]
  \centering
  \includegraphics[width=\columnwidth]{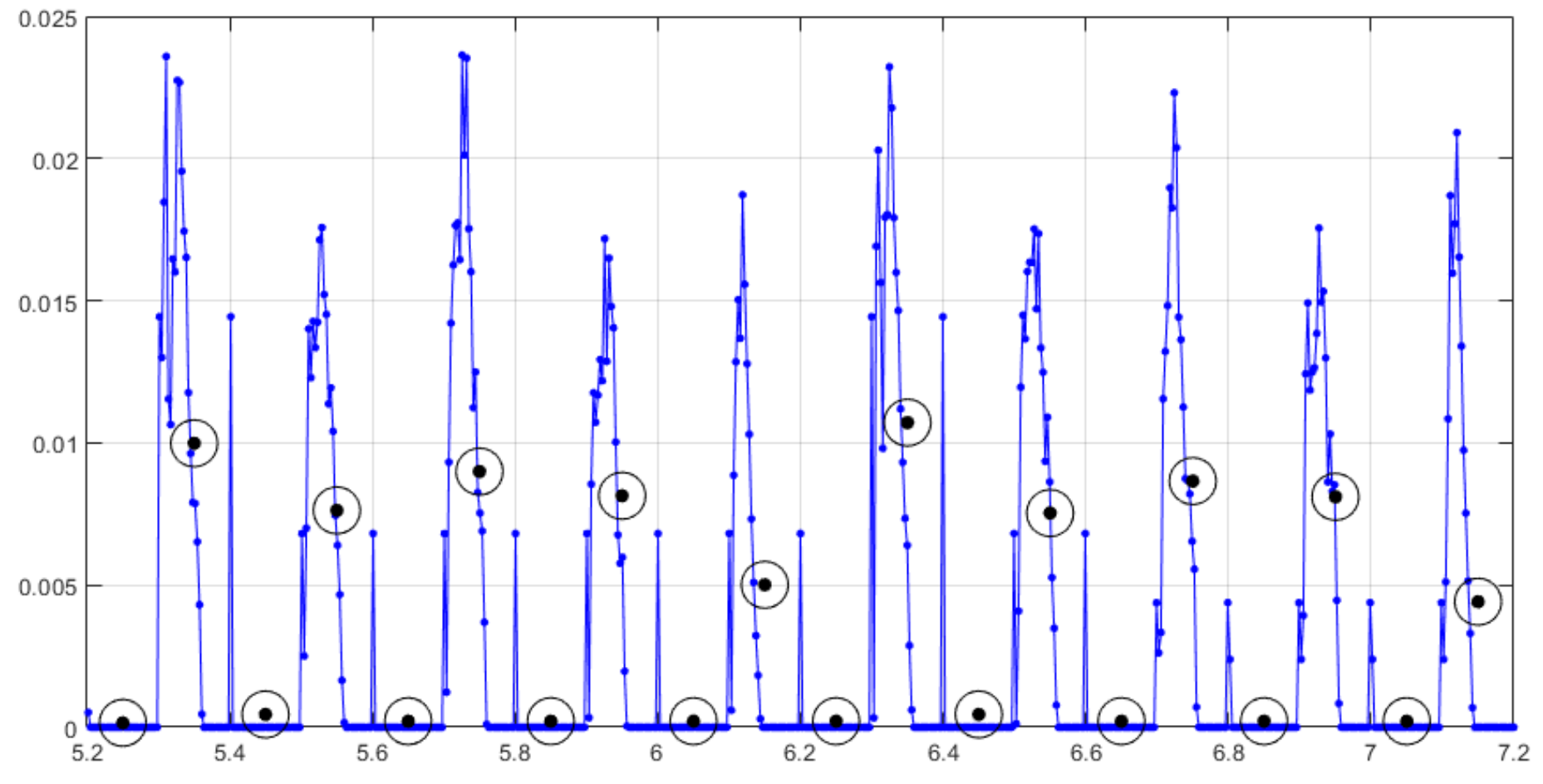}
  \caption{Comparison of a power trace at 64 frames per cycle to a power trace at 2 frames per cycle. At lower frame counts, the power sample converges to average power, and the frame size increases.}
  \label{fig:pico2frame}
\end{figure}

We also performed an ACA analysis at a frame size of 2 frames per cycle rather than 64 frames per cycle. \autoref{fig:pico2frame} shows the effect on the power trace. At wide frame size, the power converges to the average of the smaller frame size. In our experiments, we found that a wide frame size is less precise to pick out leaky gates. For the same trace set as the previous experiment, the 2-frame-per-cycle version flags only 122 cells (as opposed to 412 cells) as leaky. 

However, the use of a wider frame size may still have advantages. The 122 cells that are found at a wider frame size are a subset of the 412 cells found at a smaller frame size, with an exception of a single cell. Furthermore, there is a significant performance gain in power simulation time at wider frame sizes (\autoref{sec:acaperf}). We can thus think of power simulations at wide frame sizes as a quick assessment to determine the \ac{lti} and to scan the overall properties of side-channel leakage in the design.

\vspace{-4mm}
\subsection{Leaky Gate analysis}

\label{sec:lga1}

We analyzed the type and nature of the 412 cells that are being flagged as leaky by ACA, under the specific leakage model of the SBOX output. The direct analysis of the gate-level netlist is cumbersome because the synthesized netlist is flattened, and because most gates have non-descriptive names such as {\tt g136941}. However, it is possible to direct the synthesis tool to keep track of the originating line of RTL code that results in a specific gate. This way, we found that the 412 cells come from 47 unique sites in the RTL code. This allows the user to identify the RTL source code location of the leakage, and \autoref{tab:lgi1} summarizes the identified gates by RTL source file. 21 leaky cells are not identified by their RTL origin and are not listed in the table.

\begin{table}[t]
\centering
\caption{Leaky Gate Identification for AES Coprocessor}
\label{tab:lgi1}
\vspace{-.4cm}
\scriptsize
\begin{tabular}{llrr}\toprule
\textbf{Module} & \textbf{File} & \textbf{\# Cells} & \textbf{Sequential}\\\midrule 
Top-level & {\tt aes\_comp\_core.v}	& 5	&  0 \\
Decryption    & {\tt aes\_comp\_decipher\_block.v}	 & 29	 & 0 \\
Encryption    & {\tt aes\_comp\_encipher\_block.v}   &	204	 & 26 \\
Keymem        & {\tt aes\_comp\_key\_mem.v}          & 	1	 & 0    \\
SBOX          & {\tt comp\_sbox.v}	                 & 130	 & 0 \\
Bus Interface & {\tt picoaes.v}	                     & 22	 & 10 \\
\bottomrule
\end{tabular}
\end{table}

The list of cells in \autoref{tab:lgi1} is intriguing. ACA is able to identify non-trivial leakage, often occurring as a result of the integration of cryptographic functions. The following example illustrates this point. In the Bus interface, the IV register is flagged as a source of leakage, which is unexpected. However, upon inspection of the code, it can be shown that the IV register senses every output value of the encryption module. Furthermore, due to the sequential nature of the computation, the encryption module reflects intermediate round values, {\em including} each individual SBOX output. The results in SBOX-related leakage appear at the IV register. The identification of individual RTL files and line numbers as leaky, based on a gate-level simulation, is an important debugging tool in the hands of the designer.

\vspace{-4mm}
\subsection{Non-specific ACA}

We illustrate how ACA identifies leaky cells with a non-specific leakage model. In the following example, we create a non-specific test on the state variable in round 6 of the encryption. We use two sets of test vectors, and both contain random plaintext and key values. However, the second group contains specially selected (plaintext, key) pairs that create an all-zero round-6 state variable. Such pairs are easy to create: select a random key, and decrypt an all-zero state starting at round 4 of the decryption. The resulting plaintext is the sought starting value.

\begin{figure}[t]
  \centering
  \includegraphics[width=0.8\columnwidth]{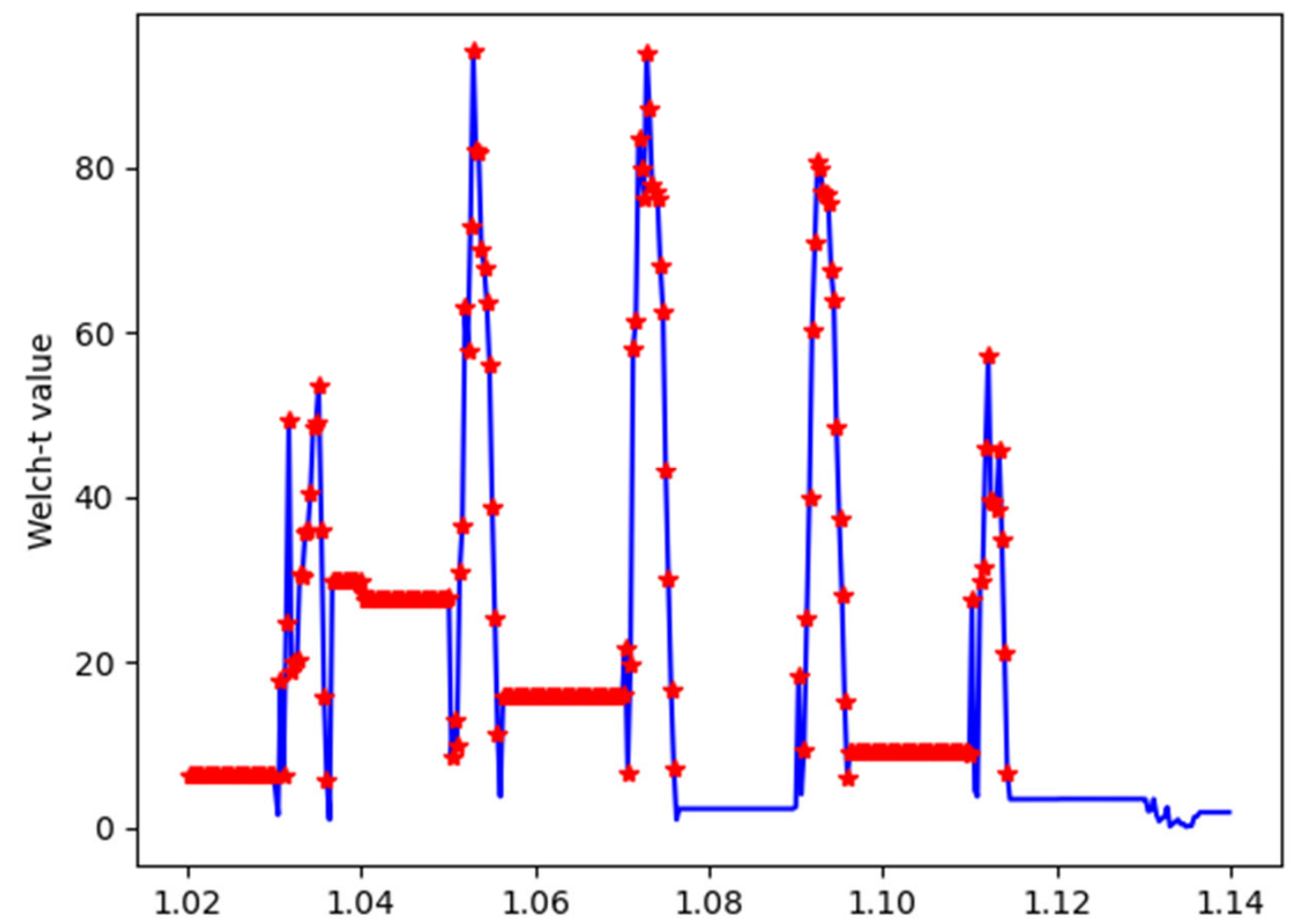}
  \caption{Leaky frames in round 6 for a non-specific test on all state bits concurrently.}
  \label{fig:nonspec2}
\end{figure}

\begin{table}[t]
\centering
\caption{Leaky Gate Identification using non-specific round-6 state bias}
\label{tab:lgi3}
\vspace{-.4cm}
\scriptsize
\begin{tabular}{llrr}\toprule
\textbf{Module} & \textbf{File} & \textbf{\# Cells} & \textbf{Sequential}\\\midrule 
Top-level & {\tt aes\_comp\_core.v}	                 & 79	&  0 \\
Decryption    & {\tt aes\_comp\_decipher\_block.v}	 & 	351   & 0 \\
Encryption    & {\tt aes\_comp\_encipher\_block.v}   &	1172  &  128\\
Keymem        & {\tt aes\_comp\_key\_mem.v}          & 	0	  & 0    \\
SBOX          & {\tt comp\_sbox.v}	                 & 	870   & 0 \\
Bus Interface & {\tt picoaes.v}	                     & 	277   & 0  \\
\bottomrule
\end{tabular}
\end{table}

Using non-specific ACA we can identify the gates that are most affected by this bias, and thus the gates that are responsible for side-channel leakage. \autoref{fig:nonspec2} illustrates the \ac{lti} on round 6, where the bias occurs. The majority of the frames (251 out of 385 in the trace) are flagged as part of LTI. Furthermore, after ranking the cells, we identify 2,812 unique cells as correlated with the round-6 state bias. This is much more than the 412 cells selected using a specific leakage model on the first key byte. However, this result is not unexpected: zero-forcing an entire state word (128 bits) where the expected value would be random is a very significant bias, which has an impact throughout the datapath. \autoref{tab:lgi3} shows the distribution of the 2,812 cells over the design. 

\begin{figure}
    \centering
    \includegraphics[width=\linewidth]{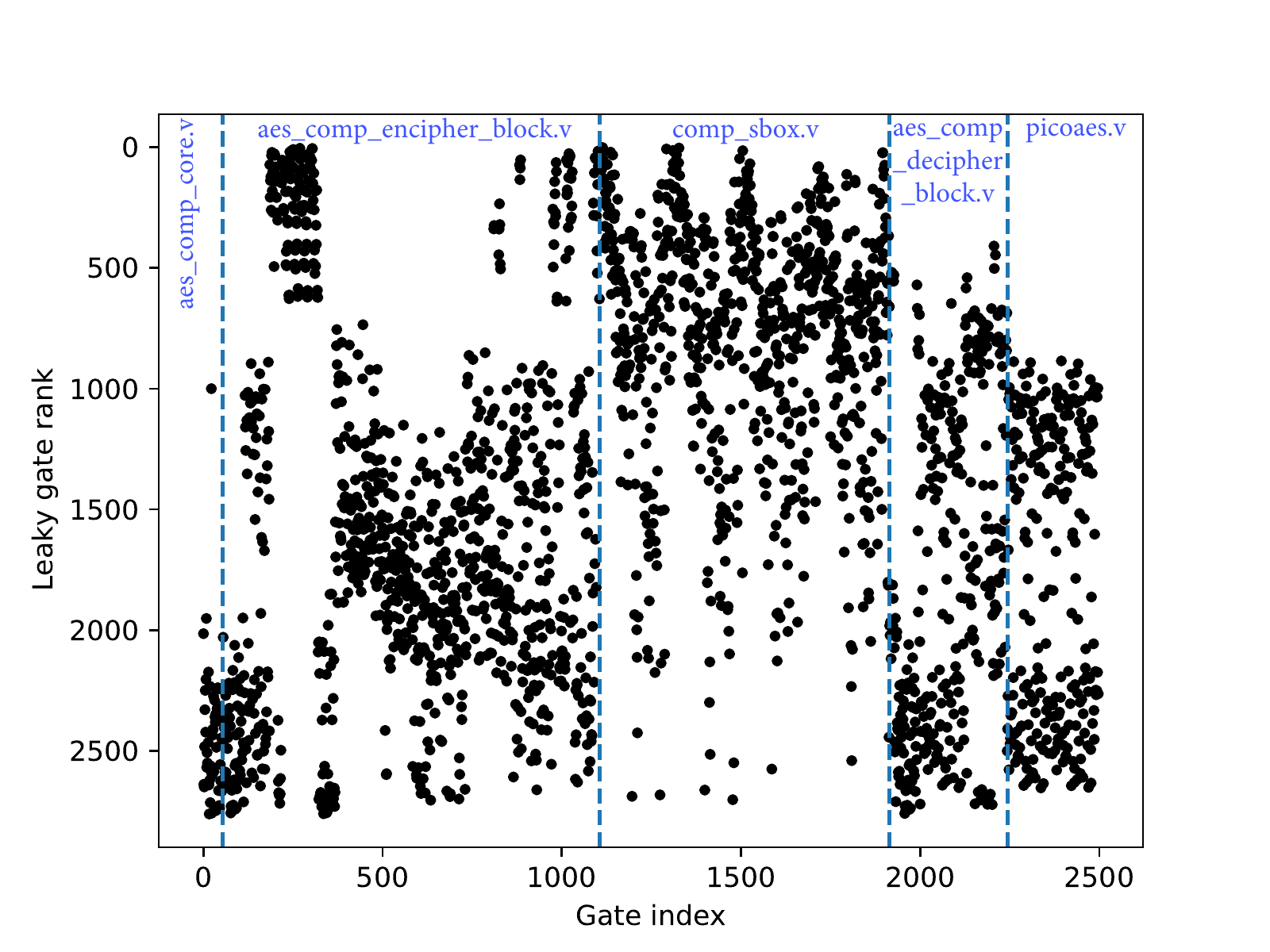}
    \vspace{-.4cm}
    \caption{Leaky gate ranks identified by non-specific test in ACA on AES coprocessor sectioned into RTL design files.}
    \label{fig:nonspec_graph_rank}
\end{figure}
Among the flagged leaky cells 2,498 gates were traced back to their RTL design files. \autoref{fig:nonspec_graph_rank} illustrates the rank of flagged leaky gates from each design file with rank=1 belonging to the leakiest gate in the design.

We caution that a strongly biased test, such as this all-zero round-6 non-specific test, always results in aggressive leaky cell selection. However, a weaker form of the test is easy to define, for example by biasing only a single state byte of round 6. The non-specific ACA test lets a user evaluate the impact of an arbitrarily chosen bias in the cipher.

\vspace{-4mm}
\section{ACA on RISC-V based SoC}

\begin{figure}[t]
  \centering
  \includegraphics[width=0.8\columnwidth]{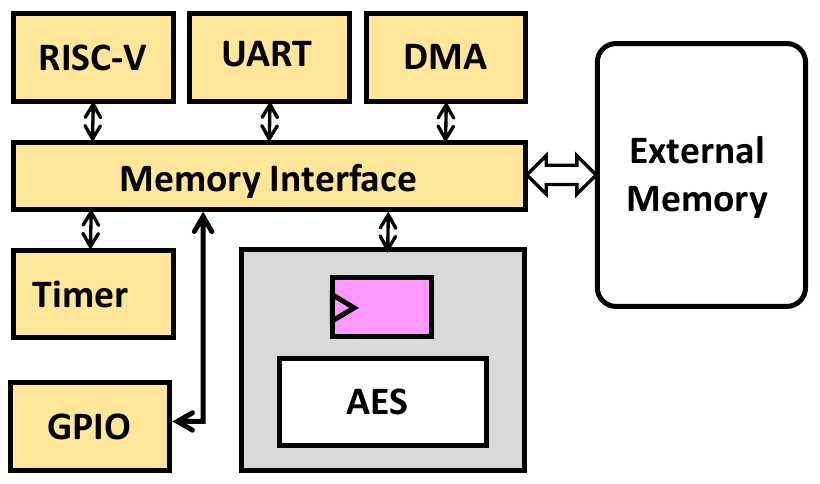}
  \caption{Block diagram of the RISC-V based SoC including the AES coprocessor}
  \label{fig:soc}
\end{figure}

\begin{table}[t]
\centering
\caption{Cell type and area for RISC-V based SoC}
\label{tab:skiva}
\vspace{-.4cm}
\scriptsize
\begin{tabular}{lrr}\toprule
\textbf{Type} & \textbf{Cell Count} & \textbf{Area (\%)} \\\midrule 
Sequential         & 8,091  & 51.5 \\
Logic              & 21,484 & 48.5 \\
Total              & 29,575 & 100.0  \\
\bottomrule
\end{tabular}
\vspace{-.5cm}
\end{table}

To investigate the scalability of ACA we also applied the methodology on the SoC shown in \autoref{fig:soc}. A 5-stage pipelined RISC-V core \color{black}(fetch, decode, execute, memory, write-back) \color{black} integrates a collection of peripherals including the memory-mapped AES coprocessor discussed in \autoref{sec:coproc}. In a typical access sequence, the RISC-V software uploads a key and a block of plaintext to the coprocessor, and then uses the control/status register of the coprocessor to start the encryption and monitor the completion flag. The RISC-V software then retrieves a block of ciphertext. This design is considerably more complicated than the stand-alone AES design, and covers a software and a hardware component. \autoref{tab:skiva} shows synthesis results for SkyWater 130nm standard cells at 50MHz clock. The overall design is three times larger than the AES coprocessor by itself.

\vspace{-4mm}
\subsection{Architecture Correlation Analysis}

In this test, we are investigating the hardware/software interface between the RISC-V software and the AES coprocessor. Therefore, we apply ACA with a specific leakage model using the Hamming Weight on the output of the pre-whitening round. This will enable the monitoring of any interactions between the plaintext and the key on the path from software to the hardware coprocessor. \autoref{fig:listing} shows a portion of the driver software. In a 32-bit architecture, a 128-bit block is loaded using 4 consecutive memory-mapped writes. The driver first loads 4 plaintext words, followed by 4 key words. Next, the coprocessor control register is configured to run a single block encryption. The software then goes into a polling loop waiting for the coprocessor to complete operation, about 50 clock cycles later.

\begin{figure}[t]
\caption{RISC-V driver software for AES Coprocessor}
\label{fig:listing}
\footnotesize
\begin{lstlisting}
  li      a4,8
  lw      a3,0(a4)     ; load plaintxt[0]        
  sw      a3,4(a5)     ; STALL
  lw      a3,4(a4)     ; load plaintxt[1]
  sw      a3,8(a5)     ; STALL
  lw      a3,8(a4)     ; load plaintxt[2]
  sw      a3,12(a5)    ; STALL
  lw      a4,12(a4)    ; load plaintxt[3]
  sw      a4,16(a5)    ; STALL
  li      a4,24   
  lw      a3,0(a4)     ; load key[0]         
  sw      a3,20(a5)    ; STALL
  lw      a3,4(a4)     ; load key[1]
  sw      a3,24(a5)    ; STALL
  lw      a3,8(a4)     ; load key[2]
  sw      a3,28(a5)    ; STALL
  lw      a4,12(a4)    ; load key[3]
  sw      a4,32(a5)    ; STALL
  li      a4,6   
  sw      a4,0(a5)     ; control
  li      a4,4   
  sw      a4,0(a5)     ; start
  li      a3,1
.L121:
  lw      a4,68(a5)
  bne     a4,a3,.L121
\end{lstlisting}
\hfill
\end{figure}

\begin{figure}[t]
  \centering
  \includegraphics[width=\columnwidth]{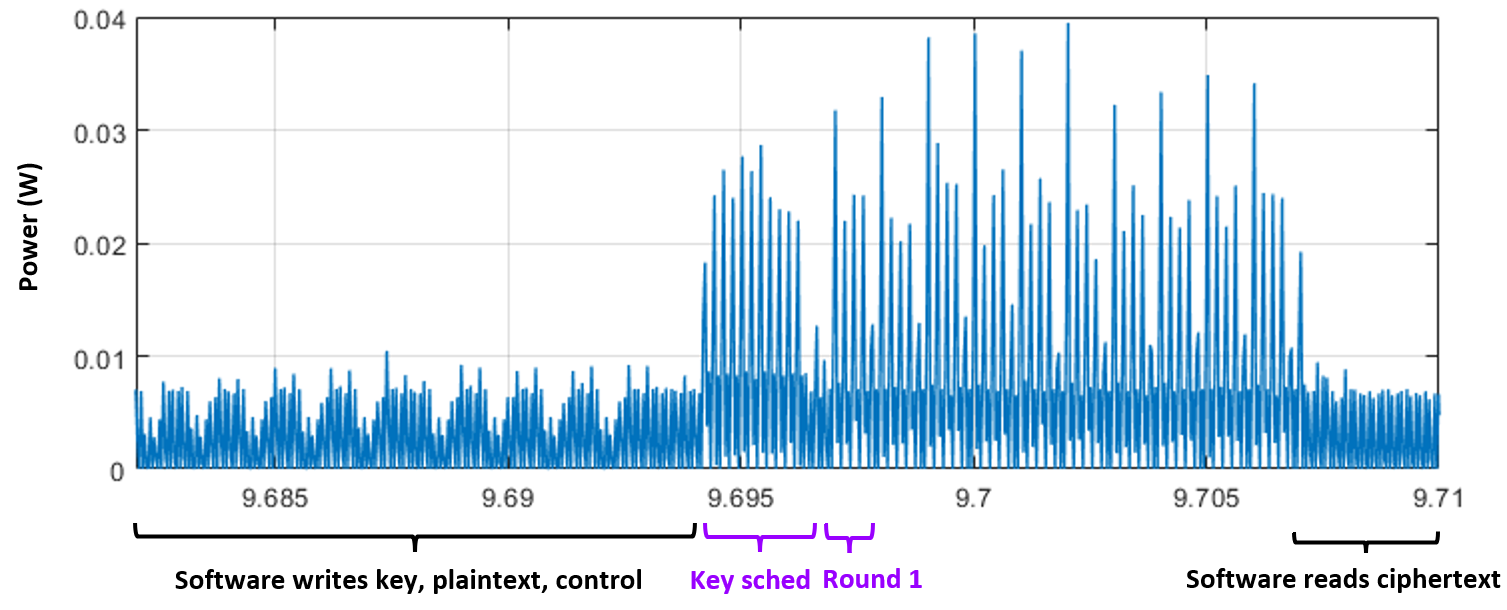}
  \vspace{-.4cm}
  \caption{Power trace of the RISC-V based SoC}
  \label{fig:socpower}
\end{figure}

\subsubsection{Results}

We selected a set of 1024 vectors under a random plaintext and a constant key. We then ran a gate-level simulation and a gate-level power simulation at 5 frames per clock cycle. 
\autoref{fig:socpower} shows a sample power trace from the simulation. The testbench covers software activity as well as hardware activity. The hardware activity uses considerably more power than software because of the higher parallelism of the hardware coprocessor implementation. However, because of the noiseless simulation, the overall operation is visible with remarkable clarity. The trace starts with the software transmitting a key value, followed by a plaintext value. \autoref{fig:socpower} shows a series of 8 notches in the power trace which correspond to reduced power consumption. These are caused by pipeline stall operations on the RISC-V processor (\autoref{fig:listing} lines 3, 5, 7, 9, 12, 14, 16, 18). Next, the software triggers the hardware AES execution, which runs the key schedule followed by 10 rounds. Finally, the RISC-V software retrieves the ciphertext.

We next ran ACA using the aforementioned specific leakage model on the Hamming Weight of the input of round 1. We identify the \ac{lti} with a $\rho_{threshold}$ of 0.2, which flags 91 frames out of the 710 frames as containing potentially leaky samples. By correlating the transitions by cells within \ac{lti} with the specific leakage model, 1,298 cells are then flagged as leaky. This group represents 4.38 \% of the total number of 29,575 cells. 

\vspace{-4mm}
\subsection{Leaky Gate Analysis}

\begin{figure}[t]
  \centering
  \includegraphics[width=\columnwidth]{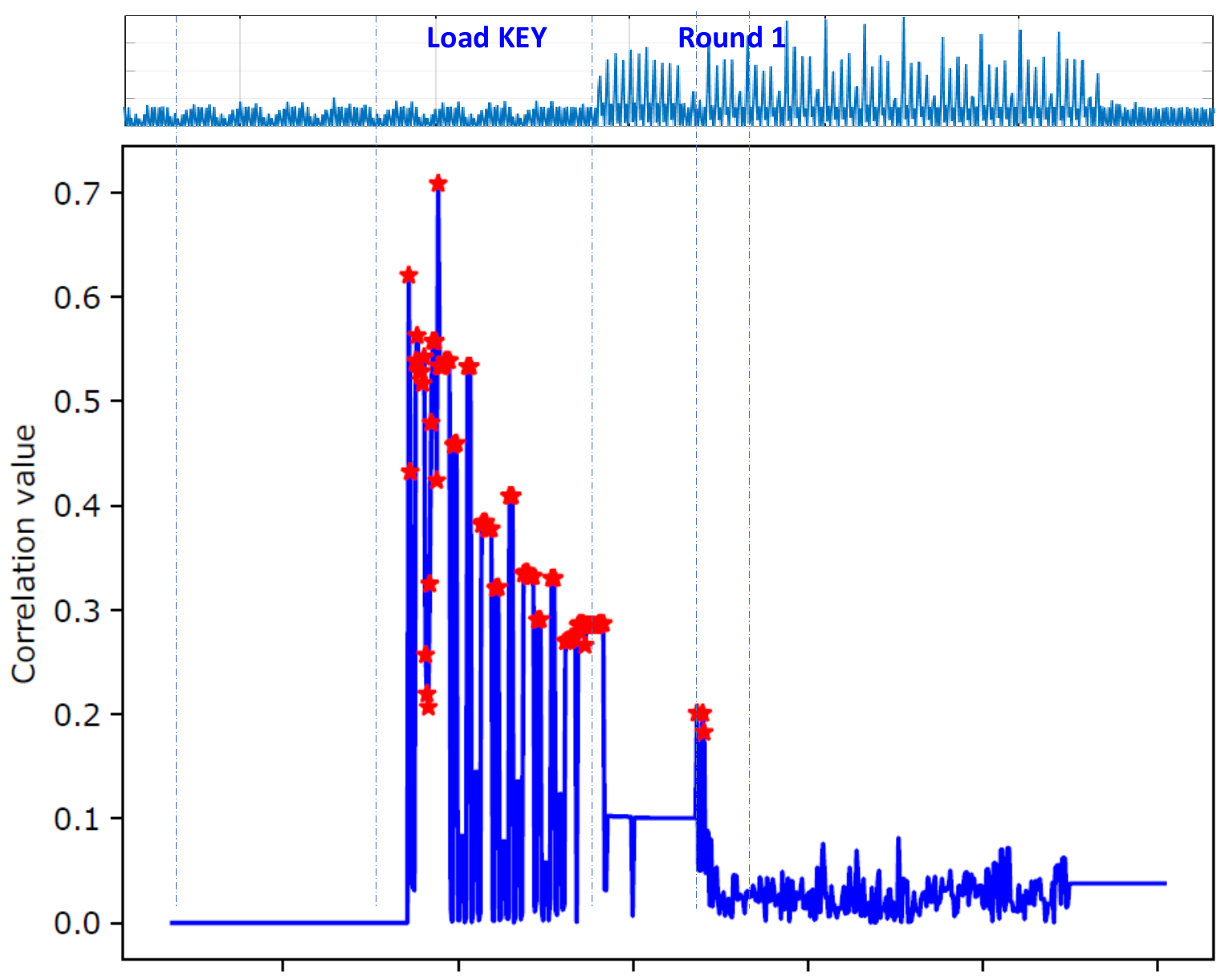}
  \caption{Leaky Frame Selection in ACA on RISC-V based SoC}
  \label{fig:skivaleaky}
\end{figure}

\autoref{fig:skivaleaky} shows the distribution of leaky frames over the testbench. The leakage model is $HW(key \oplus pt)$. Remarkably, the bulk of the leaky frames occurs during the {\em software driver} activity, while the key is being loaded into the coprocessor. In addition, there is also leakage during the first round of the encryption, which is expected. 

ACA demonstrates that the RISC-V processor, the memory bus, and the memory-mapped AES coprocessor are all potential contributors to side-channel leakage. The origin of such leakage is caused by the interaction of values over shared storage and interconnect. The leakage occurs in the processor micro-architecture because the driver software writes the key after the plaintext. There is a minor hint of this issue in the software driver itself. The first key byte is loaded in register {\tt a3}, which still contains a portion of plaintext. This leads to distance-based leakage conforming to the leakage model.

\begin{table}[t]
\centering
\caption{Leaky Gate Identification for RISC-V based SoC}
\label{tab:lgi2}
\vspace{-.4cm}
\scriptsize
\begin{tabular}{llrr}\toprule
\textbf{Module} & \textbf{File} & \textbf{\# Cells} & \textbf{Seq}\\\midrule 
\multicolumn{4}{l}{\bf AES Coprocessor} \\
AES top &  {\tt aes\_comp\_core.v}	        &    2	&0\\
Decryption &  {\tt aes\_comp\_decipher\_block.v}	&    21	&0\\
Encryption &  {\tt aes\_comp\_encipher\_block.v}	&    27	&14\\
KeyMem &  {\tt aes\_comp\_key\_mem.v	}       &     1	&0\\
Bus Interface&  {\tt aes\_top.v}	                &     130	&112\\
SBOX&  {\tt comp\_sbox.v}	            &    108	&0\\\midrule
\multicolumn{4}{l}{\bf RISC-V} \\
ALU&  {\tt ALU.v}	                    &     332	&0\\
Control&  {\tt control\_unit.v}	        	&    2	&0\\
Control&  {\tt controller.sv}	            &      10	&0\\
Memory&  {\tt memory\_arbiter.v}	        &      11	&0\\
Memory&  {\tt memory\_interface.v}	    	&    3	&0\\
Pipeline&  {\tt pipeline\_register.v}	    &    66	&31\\
Regfile&  {\tt regFile.v}	                &    248&	248\\\midrule
\multicolumn{4}{l}{\bf Peripherals} \\
GPIO&  {\tt gpio\_top.v}	            	&    3	&0\\
DMA Bus Control&  {\tt s\_axi\_controller.sv}	    &    3	&0\\
UART&  {\tt simpleuart.v}	            &    3	&0\\
DMA&  {\tt transposer.sv}	            &    3	&0\\
DMA&  {\tt ca\_prng.v}                  &    4	&0\\
DMA&  {\tt dma\_top.v}                  &    6	&0\\
DMA&  {\tt fifo\_dma.sv}	            &    3	&0\\
DMA&  {\tt fifo.v}	                	&    2	&0\\
DMA&  {\tt tDMA.sv}	                &    2	&0\\
\bottomrule
\end{tabular}
\end{table}

\autoref{tab:lgi2} shows the distribution of the 1,298 leaky cells over the design. There are indeed a large number of leaky gates located within the RISC-V processor. We analyze two examples below.

\color{black}
The highest-ranked leaky gate (with a correlation of 0.463) in the SoC is the pipeline register of the memory stage which, in its data-path components, transfers the contents of the second source register and the result of the ALU from the execute to the memory stage. Even for instructions that do not need ALU operation, the ALU result is written with the addition of the two source operands. Therefore the same transitional leakage discussed for {\tt a3} register can occur for the ALU result register. 

The leakage from the peripherals is caused by the transmission of key and plaintext on the memory interface. At the connection point of each peripheral module to the memory interface there are multiplexers to decide whether the transmitted data should be admitted to the current peripheral. Such interconnect logic can manifest Hamming weight leakage of the plaintext and key.
\color{black}

These examples demonstrate that ACA is a powerful debugging tool, as it can highlight side-channel leakage of the gate-level implementation at the RTL level. 

\vspace{-4mm}
\section{ACA Performance Considerations}

\label{sec:acaperf}

\begin{table}[t]
\centering
\caption{ACA Performance for various steps in the flow. Performance
numbers in user seconds$^*$ for 1024 Vectors.}
\label{tab:acaperf}
\vspace{-.4cm}
\scriptsize
\begin{tabular}{lrr}\toprule
                         & \textbf{AES Coprocessor} & \textbf{RISC-V based SoC} \\\midrule 
Gate-level Synthesis      & 392                      & 1,201            \\
Simulation                & 2,436                    & 6,996            \\
Power Estimation          &                    &                   \\
~~~~~64 frames/cycle      & 7,862                    &                   \\
~~~~~2 frames/cycle       & 1,557                    &                   \\
~~~~~5 frames/cycle       &                          & 31,201           \\
Correlation Analysis      & $<$60                   & $<$60              \\
\\\bottomrule
\multicolumn{3}{l}{$^*$Xeon Gold 6248 CPU @ 2.50GHz, 384G Workstation}
\end{tabular}
\vspace{-.5cm}
\end{table}

ACA adds a new design step to the overall design flow, and thus the cost of running ACA in comparison to other tools in the design flow must be considered.
\autoref{tab:acaperf} summarizes the runtime performance of ACA analysis.
There are three major components that consume the bulk of the execution time: logic synthesis, functional gate-level simulation, and gate-level power simulation. The ACA correlation component is minor and typically takes less than a minute to complete.
Overall, we observe that gate-level power simulation is a dominant factor that is more complex than gate-level synthesis and gate-level simulation. The overall runtime is strongly affected by the design size and the total number of frames per trace. On the plus side, the power simulation step is embarrassingly parallel. Each test vector can be run independently from the other. In our experiments, we did not use any parallel execution.

\vspace{-5mm}
\section{Conclusions}

Gate-level leakage assessment is a tool that supports a designer to identify leaky gates in a pre-silicon design context. Our methodology relies on industry-standard tools including logic synthesis, gate-level simulation, and gate-level power estimation, together with scripting on the intermediate results. Architecture Correlation Analysis, the underlying detection technique to support gate-level leakage assessment, can serve as a verification technique as well as as a basis for countermeasure design. In particular, by moving the leaky gates flagged by ACA into a separate power domain, a low-cost countermeasure may be enabled that requires only selective replacement of cells in a design.

\IEEEdisplaynontitleabstractindextext
\IEEEpeerreviewmaketitle

\bibliographystyle{IEEEtran}
\bibliography{Bibliography,SoK}

\begin{thebibliography}{10}
\providecommand{\url}[1]{#1}
\csname url@samestyle\endcsname
\providecommand{\newblock}{\relax}
\providecommand{\bibinfo}[2]{#2}
\providecommand{\BIBentrySTDinterwordspacing}{\spaceskip=0pt\relax}
\providecommand{\BIBentryALTinterwordstretchfactor}{4}
\providecommand{\BIBentryALTinterwordspacing}{\spaceskip=\fontdimen2\font plus
\BIBentryALTinterwordstretchfactor\fontdimen3\font minus
  \fontdimen4\font\relax}
\providecommand{\BIBforeignlanguage}[2]{{%
\expandafter\ifx\csname l@#1\endcsname\relax
\typeout{** WARNING: IEEEtran.bst: No hyphenation pattern has been}%
\typeout{** loaded for the language `#1'. Using the pattern for}%
\typeout{** the default language instead.}%
\else
\language=\csname l@#1\endcsname
\fi
#2}}
\providecommand{\BIBdecl}{\relax}
\BIBdecl

\bibitem{nikova2006threshold}
S.~Nikova, C.~Rechberger, and V.~Rijmen, ``Threshold implementations against
  side-channel attacks and glitches,'' in \emph{International conference on
  information and communications security}.\hskip 1em plus 0.5em minus
  0.4em\relax Springer, 2006, pp. 529--545.

\bibitem{DBLP:conf/ches/Moradi14}
\BIBentryALTinterwordspacing
A.~Moradi, ``Side-channel leakage through static power - should we care about
  in practice?'' in \emph{Cryptographic Hardware and Embedded Systems - {CHES}
  2014 - 16th International Workshop, Busan, South Korea, September 23-26,
  2014. Proceedings}, ser. Lecture Notes in Computer Science, L.~Batina and
  M.~Robshaw, Eds., vol. 8731.\hskip 1em plus 0.5em minus 0.4em\relax Springer,
  2014, pp. 562--579. [Online]. Available:
  \url{https://doi.org/10.1007/978-3-662-44709-3\_31}
\BIBentrySTDinterwordspacing

\bibitem{DBLP:conf/sersc-isa/ChenHS09}
\BIBentryALTinterwordspacing
Z.~Chen, S.~Haider, and P.~Schaumont, ``Side-channel leakage in masked circuits
  caused by higher-order circuit effects,'' in \emph{Advances in Information
  Security and Assurance, Third International Conference and Workshops, {ISA}
  2009, Seoul, Korea, June 25-27, 2009. Proceedings}, ser. Lecture Notes in
  Computer Science, J.~H. Park, H.~Chen, M.~Atiquzzaman, C.~Lee, T.~Kim, and
  S.~Yeo, Eds., vol. 5576.\hskip 1em plus 0.5em minus 0.4em\relax Springer,
  2009, pp. 327--336. [Online]. Available:
  \url{https://doi.org/10.1007/978-3-642-02617-1\_34}
\BIBentrySTDinterwordspacing

\bibitem{10.1145/3322483}
\BIBentryALTinterwordspacing
I.~Giechaskiel, K.~Eguro, and K.~B. Rasmussen, ``Leakier wires: Exploiting fpga
  long wires for covert- and side-channel attacks,'' \emph{ACM Trans.
  Reconfigurable Technol. Syst.}, vol.~12, no.~3, aug 2019. [Online].
  Available: \url{https://doi.org/10.1145/3322483}
\BIBentrySTDinterwordspacing

\bibitem{yao2020}
Y.~Yao, T.~Kathuria, B.~Ege, and P.~Schaumont, ``Architecture correlation
  analysis (aca): identifying the source of side-channel leakage at
  gate-level,'' in \emph{2020 IEEE International Symposium on Hardware Oriented
  Security and Trust (HOST)}.\hskip 1em plus 0.5em minus 0.4em\relax IEEE,
  2020, pp. 188--196.

\bibitem{yao2021pre}
Y.~Yao, T.~Tufan, T.~Kathuria, B.~Ege, U.~Guler, and P.~Schaumont,
  ``Pre-silicon architecture correlation analysis (paca): Identifying and
  mitigating the source of side-channel leakage at gate-level.'' \emph{IACR
  Cryptol. ePrint Arch.}, vol. 2021, p. 530, 2021.

\bibitem{cryptoeprint:2021:497}
I.~Buhan, L.~Batina, Y.~Yarom, and P.~Schaumont, ``Sok: Design tools for
  side-channel-aware implementations,'' Cryptology ePrint Archive, Report
  2021/497, 2021, \url{https://ia.cr/2021/497}.

\bibitem{PINPAS_2003}
J.~den Hartog, J.~Verschuren, E.~P. de~Vink, J.~de~Vos, and W.~Wiersma,
  ``{PINPAS:} a tool for power analysis of smartcards,'' in \emph{{SEC}}, 2003,
  pp. 453--457.

\bibitem{ELMO_2017}
D.~McCann, E.~Oswald, and C.~Whitnall, ``Towards practical tools for side
  channel aware software engineering: `grey box' modelling for instruction
  leakages,'' in \emph{{USENIX} Security Symposium}, 2017, pp. 199--216.

\bibitem{ROSITA_2019}
M.~A. Shelton, N.~Samwel, L.~Batina, F.~Regazzoni, M.~Wagner, and Y.~Yarom,
  ``Rosita: Towards automatic elimination of power-analysis leakage in
  ciphers,'' in \emph{NDSS}, 2021.

\bibitem{EMSIM_2020}
N.~Sehatbakhsh, B.~B. Yilmaz, A.~G. Zajic, and M.~Prvulovic, ``{EMSim}: A
  microarchitecture-level simulation tool for modeling electromagnetic
  side-channel signals,'' in \emph{{HPCA}}, 2020, pp. 71--85.

\bibitem{CASCADE_2020}
D.~Sijacic, J.~Balasch, B.~Yang, S.~Ghosh, and I.~Verbauwhede, ``Towards
  efficient and automated side-channel evaluations at design time,'' \emph{J.
  Cryptogr. Eng.}, vol.~10, no.~4, pp. 305--319, 2020.

\bibitem{10.1145/3383445}
\BIBentryALTinterwordspacing
A.~Nahiyan, J.~Park, M.~He, Y.~Iskander, F.~Farahmandi, D.~Forte, and
  M.~Tehranipoor, ``Script: A cad framework for power side-channel
  vulnerability assessment using information flow tracking and pattern
  generation,'' \emph{ACM Trans. Des. Autom. Electron. Syst.}, vol.~25, no.~3,
  may 2020. [Online]. Available: \url{https://doi.org/10.1145/3383445}
\BIBentrySTDinterwordspacing

\bibitem{DBLP:conf/samos/RegazzoniBEGPDMPPLI07}
\BIBentryALTinterwordspacing
F.~Regazzoni, S.~Badel, T.~Eisenbarth, J.~Gro{\ss}sch{\"{a}}dl, A.~Poschmann,
  Z.~T. Deniz, M.~Macchetti, L.~Pozzi, C.~Paar, Y.~Leblebici, and P.~Ienne, ``A
  simulation-based methodology for evaluating the dpa-resistance of
  cryptographic functional units with application to {CMOS} and {MCML}
  technologies,'' in \emph{Proceedings of the 2007 International Conference on
  Embedded Computer Systems: Architectures, Modeling and Simulation {(IC-SAMOS}
  2007), Samos, Greece, July 16-19, 2007}, 2007, pp. 209--214. [Online].
  Available: \url{https://doi.org/10.1109/ICSAMOS.2007.4285753}
\BIBentrySTDinterwordspacing

\bibitem{karna2019}
P.~SLPSK, P.~K. Vairam, C.~Rebeiro, and K.~Veezhinathan, ``Karna: A gate-sizing
  based security aware eda flow for improved power side-channel attack
  protection,'' in \emph{Proceedings of the International Conference on
  Computer-Aided Design, {ICCAD} 2019, Westminster, CO, USA, November 04-07,
  2019}.

\bibitem{RTL_PSC_2019}
M.~T. He, J.~Park, A.~Nahiyan, A.~Vassilev, Y.~Jin, and M.~M. Tehranipoor,
  ``{RTL-PSC:} automated power side-channel leakage assessment at
  register-transfer level,'' in \emph{{VTS}}, 2019, pp. 1--6.

\bibitem{DBLP:conf/host/FGBR20}
\BIBentryALTinterwordspacing
M.~A.~K. F, V.~Ganesan, R.~Bodduna, and C.~Rebeiro, ``{PARAM:} {A}
  microprocessor hardened for power side-channel attack resistance,'' in
  \emph{2020 {IEEE} International Symposium on Hardware Oriented Security and
  Trust, {HOST} 2020, San Jose, CA, USA, December 7-11, 2020}.\hskip 1em plus
  0.5em minus 0.4em\relax {IEEE}, 2020, pp. 23--34. [Online]. Available:
  \url{https://doi.org/10.1109/HOST45689.2020.9300263}
\BIBentrySTDinterwordspacing

\bibitem{DBLP:conf/uss/GigerlHPMB21}
\BIBentryALTinterwordspacing
B.~Gigerl, V.~Hadzic, R.~Primas, S.~Mangard, and R.~Bloem, ``Coco: Co-design
  and co-verification of masked software implementations on cpus,'' in
  \emph{30th {USENIX} Security Symposium, {USENIX} Security 2021, August 11-13,
  2021}, M.~Bailey and R.~Greenstadt, Eds.\hskip 1em plus 0.5em minus
  0.4em\relax {USENIX} Association, 2021, pp. 1469--1468. [Online]. Available:
  \url{https://www.usenix.org/conference/usenixsecurity21/presentation/gigerl}
\BIBentrySTDinterwordspacing

\bibitem{gilbert2011testing}
G.~Goodwill, B.~Jun, J.~Jaffe, P.~Rohatgi \emph{et~al.}, ``A testing
  methodology for side-channel resistance validation,'' in \emph{NIST
  non-invasive attack testing workshop}, vol.~7, 2011, pp. 115--136.

\bibitem{cryptoeprint:2021:1235}
P.~Kiaei, Z.~Liu, R.~K. Eren, Y.~Yao, and P.~Schaumont, ``Saidoyoki: Evaluating
  side-channel leakage in pre- and post-silicon setting,'' Cryptology ePrint
  Archive, Report 2021/1235, 2021, \url{https://ia.cr/2021/1235}.

\bibitem{kiaei2022leverage}
P.~Kiaei, Z.~Liu, and P.~Schaumont, ``Leverage the average: Averaged sampling
  in pre-silicon side-channel leakage assessment,'' in \emph{Proceedings of the
  2022 on Great Lakes Symposium on VLSI}, 2022.

\bibitem{DBLP:conf/host/DasNCGS19}
\BIBentryALTinterwordspacing
D.~Das, M.~Nath, B.~Chatterjee, S.~Ghosh, and S.~Sen, ``{STELLAR:} {A} generic
  {EM} side-channel attack protection through ground-up root-cause analysis,''
  in \emph{{IEEE} International Symposium on Hardware Oriented Security and
  Trust, {HOST} 2019, McLean, VA, USA, May 5-10, 2019}.\hskip 1em plus 0.5em
  minus 0.4em\relax {IEEE}, 2019, pp. 11--20. [Online]. Available:
  \url{https://doi.org/10.1109/HST.2019.8740839}
\BIBentrySTDinterwordspacing

\end{thebibliography}

\begin{IEEEbiographynophoto}{Pantea Kiaei} (Student Member, IEEE) is a Ph.D. student in Electrical and Computer Engineering at Worcester Polytechnic Institute. She received her MS degree in Computer Engineering from Virginia Tech in 2019 and prior to that received her BS degree in Electrical Engineering from Sharif University of Technology, Iran, in 2017. She has reviewed papers for ACM TECS, ACM JETC, and IEEE TVLSI journals. Her research interests include secure hardware design, computer architecture, and trustworthy secure systems.
\end{IEEEbiographynophoto}

\begin{IEEEbiographynophoto}{Yuan Yao} received her bachelor's degree in Electronic Engineering from Northwestern Polytechnical University, Xi'an, China in 2014. She got her Master's Degree in Electrical and Computer Engineering from Cornell University, Ithaca, US in 2016. Currently she is a Ph.D. candidate at the Bradley Department of Electrical and Computer Engineering, Virginia Tech. She serves as reviewer for several IEEE and ACM journals. Her research area include pre-silicon side-channel analysis, side-channel attacks and countermeasures, fault attacks and countermeasures, secure hardware design. 
\end{IEEEbiographynophoto} 

\begin{IEEEbiographynophoto}{Zhenyuan Liu} (Student Member, IEEE) is a Ph.D. student in Electrical and Computer Engineering at Worcester Polytechnic Institute. She received her MS degree in Electrical and Computer Engineering from Worcester Polytechnic Institute in 2020 and prior to that received her BS degree in Engineering of Science from Trinity University, San Antonio, Texas, in 2019. Her research interests include side-channel attacks, leakage assessments and micro-architectural hardware security.
\end{IEEEbiographynophoto}

\begin{IEEEbiographynophoto}{Nicole Fern} is a Senior Security Analyst at Riscure. She received her undergraduate degree in Electrical Engineering from The Cooper Union for the Advancement of Science and Art (2011) and her PhD degree in Electrical \& Computer Engineering from University of California, Santa Barbara (2016). She continued her research in hardware security as a post-doc before joining industry in 2018. She previously worked at Tortuga Logic, a hardware security startup, for 2.5 years before joining Riscure in June of 2021.
\end{IEEEbiographynophoto}

\begin{IEEEbiographynophoto}{Cees-Bart Breunesse} is a principal security analyst at Riscure North America. He received his Ph.D. in Computer Science from Radboud University, Netherlands, in 2005. Prior to that, he received his MSc degree in Computer Science in 2000 from Utrecht University, Netherlands. His research interests include security of embedded devices, side-channel, and fault injection.
\end{IEEEbiographynophoto}

\begin{IEEEbiographynophoto}{Jasper van Woudenberg} (@jzvw) currently is CTO for Riscure North America and half of the authors of the "Hardware Hacking Handbook: Breaking Embedded Security with Hardware Attacks". He works with Riscure's San Francisco based team to improve embedded device security through innovation.

In the past, Jasper worked for a penetration testing firm, where he performed source code review, binary reverse engineering and tested application and network security. 

At Riscure, Jasper's expertise has grown to include various aspects of hardware security; from design review and logical testing, to side-channel analysis and perturbation attacks.

Jasper has spoken at many security conferences including BlackHat briefings and trainings, Intel Security Conference, RWC, RSA, EDSC, BSides SF, Shakacon, ICMC, Infiltrate, has presented scientific research at SAC, WISSEC, CT-RSA, FDTC, ESC Design {West,East}, ARM TechCon, has reviewed papers for CHES and JC(rypto)EN, and has given invited talks at Stanford, NPS, GMU and the University of Amsterdam.
\end{IEEEbiographynophoto}

\begin{IEEEbiographynophoto}{Kate Gillis} received a B.S. in Electrical and Computer Engineering from Worcester Polytechnic Institute (WPI) in 2016. She has been at Intrinsix Corp. in Marlborough, MA since the same year, where she is currently a Verification Engineer and has worked to design and implement block-level, SoC and mixed signal verification on a variety of ASIC designs. Her awards and honors include the Provost's MQP (senior capstone) Award and the Salisbury Prize, both at WPI.
\end{IEEEbiographynophoto}

\begin{IEEEbiographynophoto}{Alex Dich} received a B.S. in electrical and computer engineering from Worcester Polytechnic Institute, Worcester, MA in 2014. He has been at Intrinsix Corp., Marlborough, MA since 2014, where he is a Senior Design Engineer focused on front-end digital ASIC design. His interests include digital signal processing, cryptography, hardware acceleration, and processor architecture.
\end{IEEEbiographynophoto}

\begin{IEEEbiographynophoto}{Peter J. Grossmann} received a B.S. in engineering from Harvey Mudd College, Claremont, CA in 2001, an M.S. in electrical engineering from the University of Washington, Seattle, WA in 2006, and a PhD in computer engineering from Northeastern University, Boston, MA in 2013.  He has been at Intrinsix Corp., Marlborough, MA since 2019, where he is currently a Solutions Architect focused on advanced research.  He worked as Associate Staff and Technical Staff at MIT Lincoln Laboratory from 2007 to 2019, and as an ASIC Design Engineer at Zilog, Inc. from 2001 to 2004.  His research interests include architecture and CAD for field programmable gate arrays, low power digital circuit design, and enhancing electronic design automation flows for security.  Dr. Grossmann’s awards and honors include the 2017 MIT Lincoln Laboratory Best Invention award and the University of Washington Top Scholar award.
\end{IEEEbiographynophoto}

\begin{IEEEbiographynophoto}{Patrick Schaumont} (Senior Member, IEEE) is a Professor in Computer Engineering at WPI. He received the Ph.D. degree in Electrical Engineering from UCLA in 2004 and the MS degree in Computer Science from Ghent University in 1990. He was a staff researcher at IMEC, Belgium from 1992 to 2000. He was a faculty member with Virginia Tech from 2005 to 2019. He joined WPI in 2020. He was a visiting researcher at the National Institute of Information and Telecommunications Technology (NICT), Japan in 2014. He was a visiting researcher at Laboratoire d'Informatique de Paris 6 in Paris, France in 2018. He is a Radboud Excellence Initiative Visiting Faculty with Radboud University, Netherlands from 2020. His research interests are in design and design methods of secure, efficient and real-time embedded computing systems. 
\end{IEEEbiographynophoto}

\end{document}